\documentclass[a4paper,english,10pt]{article}

\usepackage[utf8]{inputenc}
\usepackage[a4paper]{geometry}
\usepackage{cite}
\usepackage[inline]{enumitem}
\usepackage{verbatim,fancyvrb}
\usepackage{hyperref}

\usepackage{amsmath,amsthm,amsfonts,amssymb,stmaryrd,mathtools,relsize}
\usepackage{etex,etoolbox,thmtools,environ}
\usepackage{xspace}
\usepackage{listings}
\usepackage{tikz,graphicx,placeins}
\usetikzlibrary{calc,arrows,positioning,intersections,decorations.pathreplacing}

\newif\ifdraft
\newif\ifappendix

\appendixtrue

\newcommand{\suppressed}[1]{
	\PackageWarning{}{#1 suppressed on page \thepage\space on input line \number\inputlineno!}}

\ifdraft
	\PackageWarning{}{Draft macros enabled.}
\else
	\renewcommand{\marginpar}[1]{\suppressed{Margin paragraph}}
\fi

\ifappendix
	\newcommand{\assumeapx}[1]{#1}
\else
	\newcommand{\assumeapx}[1]{\suppressed{Content assuming the Appendix}}
\fi

\ifdraft
	\usepackage[breakable]{tcolorbox}
	\newenvironment{ccbox}[2][]{
			\begin{tcolorbox}[colback=#2!20,colframe=#2,
				breakable,
				boxsep=3pt,boxrule=.5pt,#1]
			}{\end{tcolorbox}}

\else
	\usepackage{xcomment}
	\newxcomment{ccbox}
	\newxcomment{alertbox}
	\newxcomment{commentbox}
	\newxcomment{highlightbox}
	\newxcomment{draftbox}
	\newxcomment{todobox}
\fi

\theoremstyle{plain}\newtheorem{theorem}{Theorem}
\theoremstyle{plain}
\theoremstyle{plain}
\theoremstyle{plain}\newtheorem{lemma}[theorem]{Lemma}
\theoremstyle{plain}\newtheorem{definition}{Definition}
\theoremstyle{plain}\newtheorem{remark}{Remark}
\theoremstyle{plain}

\makeatletter

\ifcsname spn@wtheorem\endcsname
	\providecommand{\@lastarg}[1]{#1}
\else
	\providecommand{\@lastarg}[4]{#4}
\fi
\newcounter{freshthmlabel}

\newcommand\fixstatement[2][\proofname\space of]{%
  \ifcsname thmt@original@#2\endcsname
    \AtEndEnvironment{#2}{%
      \xdef\pat@label{\expandafter\expandafter\expandafter\@lastarg\csname thmt@original@#2\endcsname}
      \xdef\pat@proofof{\@nameuse{pat@proofof@#2}}%
      \stepcounter{freshthmlabel}
      \label{proofatend:\thefreshthmlabel}
    }%
  \else
	\ifcsname #2name\endcsname
		\AtEndEnvironment{#2}{%
			\xdef\pat@label{\expandafter\expandafter\expandafter\@lastarg\csname #2name\endcsname}
			\xdef\pat@proofof{\@nameuse{pat@proofof@#2}}%
			\stepcounter{freshthmlabel}
			\label{proofatend:\thefreshthmlabel}
		}%
	\else
		\AtEndEnvironment{#2}{%
			\xdef\pat@label{\expandafter\expandafter\expandafter\@lastarg\csname #1\endcsname}
			\xdef\pat@proofof{\@nameuse{pat@proofof@#2}}%
			\stepcounter{freshthmlabel}
			\label{proofatend:\thefreshthmlabel}
		}%
    \fi
  \fi
  \@namedef{pat@proofof@#2}{#1}%
}

\globtoksblk\prooftoks{50}
\newcounter{proofcount}

\NewEnviron{proofatend}{%
	\ifappendix
		\begin{proof}
			\ifdraft
				 See Appendix \ref{proofatend:proofs},
				 page \pageref{proofatend:proof-\thefreshthmlabel}.
			 \else
				 See Appendix \ref{proofatend:proofs}.
			 \fi
		\end{proof}
	\fi
	\edef\next{%
		\noexpand\begin{proof}[{\pat@proofof\space\pat@label\space\noexpand\ref{proofatend:\thefreshthmlabel}}]%
		\noexpand\phantomsection%
		\noexpand\label{proofatend:proof-\thefreshthmlabel}%
		\unexpanded\expandafter{\BODY}}%
		\global\toks\numexpr\prooftoks+\value{proofcount}\relax=\expandafter{\next\end{proof}}
	\stepcounter{proofcount}}
	
	\def\printproofs{%
		\ifnum0<\value{proofcount}
			\section{Omitted proofs}
			\label{proofatend:proofs}%
			\count@=\z@
			\loop
				\the\toks\numexpr\prooftoks+\count@\relax
				\ifnum\count@<\value{proofcount}%
					\advance\count@\@ne
			\repeat
		\fi
	}
\makeatother

\fixstatement{theorem}
\fixstatement{proposition}
\fixstatement{corollary}
\fixstatement{lemma}

\newlength{\wideitemsep}
\let\olditem\item
\renewcommand{\item}{\setlength{\itemsep}{\wideitemsep}\olditem}

\newcommand{\starsim}{\mathop{\stackrel{_*}{\approx}}}

\newcommand{\?}[1]{}

\makeatletter
	\newcommand{\footnoteref}[1]{%
		\protected@xdef\@thefnmark{\ref{#1}}\@footnotemark%
	}
\makeatother

\makeatletter
\newcommand{\customlabel}[2]{%
	\protected@write \@auxout {}{\string \newlabel {#1}{{#2}{\thepage}{#2}{#1}{}} }%
	\hypertarget{#1}{#2}
}
\makeatother

\newcommand{\rtlabel}[2]{
	{\text{\scriptsize\;(\customlabel{#1}{\text{\sc #2}}\!\!)}}
}

\newcommand{\tccsm}{\ensuremath{\mathit{TCCS^m}}\xspace}
\newcommand{\octm}{\ensuremath{\mathit{OCTM}}\xspace} 

\newcommand{\defeq}{\triangleq}
\newcommand{\defiff}{\stackrel{\triangle}{\iff}}

\DeclareMathOperator*{\dom}{dom}

\newcommand{\from}{\leftarrow}

\newcommand{\sqsubsetsim}{\mathrel{%
	\ooalign{\raise0.2ex\hbox{$\sqsubset$}%
	\cr\hidewidth\raise-0.8ex\hbox{\scalebox{0.9}{$\sim$}}%
	\hidewidth\cr}}
}
\newcommand{\sqsupsetsim}{\mathrel{%
	\ooalign{\raise0.2ex\hbox{$\sqsupset$}%
	\cr\hidewidth\raise-0.8ex\hbox{\scalebox{0.9}{$\sim$}}%
	\hidewidth\cr}}
}

\newcommand{\textcode}[1]{\textnormal{\texttt{#1}}}
\newcommand{\ctmretry}{\textcode{retry}}
\newcommand{\ctmabort}[1]{\textcode{abort}\; {#1}}

\newcommand{\ctmreturn}[1]{\textcode{return}\; {#1}}
\newcommand{\ctmfork}[1]{\textcode{fork}\; {#1}}
\newcommand{\ctmnew}[1]{\textcode{newVar}\; {#1}}
\newcommand{\ctmread}[1]{\textcode{readVar}\; {#1}}
\newcommand{\ctmwrite}[2]{\textcode{writeVar}\; {#1}\; {#2}}
\newcommand{\ctmatomic}[2]{\textcode{atomic}\; {#1}\; {#2}}

\newcommand{\ctmbind}{\;\textcode{>\kern-.3ex>\kern-.1ex=}\;}
\newcommand{\ctmbnd}{\;\textcode{>\kern-.3ex>}\;}
\newcommand{\ctmBind}[2]{{#1}\ctmbind {#2}}

\newcommand{\bl}{\textcode{(}}
\newcommand{\br}{\textcode{)}}
\newcommand{\void}{\textcode{()}}

\newcommand{\ctmiso}{\textcode{isolated}\;}

\newcommand{\ctmthread}[2][t]{(\mspace{-3.7mu}[#2]\mspace{-3.7mu})_{#1}}
\newcommand{\ctmthreadtr}[5]{(\mspace{-3.7mu}[{#1 \triangleright #2 ; #3}]\mspace{-3.7mu})_{#4,#5}}
\newcommand{\ctmthreadtw}[4]{(\mspace{-3.7mu}[{#1 \triangleright #2}]\mspace{-3.7mu})_{#3,#4}}

\newcommand{\threadnames}[1]{\mathop{\mathsf{threads}}(#1)}
\newcommand{\transactionnames}[1]{\mathop{\mathsf{transactions}}(#1)}

\newcommand{\abLabel}{ab}
\newcommand{\sigabLabel}{\widehat{ab}}
\newcommand{\coLabel}{co}
\newcommand{\newLabel}{new}
\newcommand{\abArrow}[2][k]{\xrightarrow{\abLabel_{#1}{#2}}}
\newcommand{\sigabArrow}[2][k]{\xrightarrow{\sigabLabel_{#1}{#2}}}
\newcommand{\coArrow}[1][k]{\xrightarrow{\coLabel_{#1}}}
\newcommand{\newArrow}[1][k]{\xrightarrow{\newLabel_{#1}}}
\newcommand{\betaArrow}{\xrightarrow{\beta}}

\newcommand{\evalTermArrow}{\to}

\newcommand{\trArrow}{\xrightarrow{\!}}

\newcommand{\ctxhole}[1][-]{[#1]}
\newcommand{\ctxP}[2][t]{\mathbb P_{#1}\ctxhole[#2]}
\newcommand{\ctxE}[1]{\mathbb E\ctxhole[#1]}
\newcommand{\ctxT}[2][t,k]{\mathbb T_{#1}\ctxhole[#2]}

\newcommand{\Loc}{\mathsf{Loc}}
\newcommand{\Var}{\mathsf{Var}}
\newcommand{\Term}{\mathsf{Term}}
\newcommand{\TrName}{\mathsf{TrName}}
\newcommand{\State}{\mathsf{State}}
\newcommand{\Proc}{\mathsf{Proc}}

\newcommand{\ctmSt}[2]{\langle #2; #1\rangle}

\newcommand{\tequiv}{\cong_t}

\newcommand{\tccssum}{\textstyle{\sum}}
\newcommand{\tccsprod}{\textstyle{\prod}}

\newcommand{\llangle}{\langle\mspace{-4mu}\langle}
\newcommand{\rrangle}{\rangle\mspace{-4mu}\rangle}
\newcommand{\llanglebf}{%
\langle\mspace{-6.8mu}\langle\mspace{-6.8mu}\langle\mspace{-6.8mu}\langle\mspace{-6.8mu}%
\langle}
\newcommand{\rranglebf}{%
\rangle\mspace{-6.8mu}\rangle\mspace{-6.8mu}\rangle\mspace{-6.8mu}\rangle\mspace{-6.8mu}%
\rangle}

\newcommand{\tccstrana}[3][k]{\llangle #2\mspace{-1mu}\triangleright_{#1}\mspace{-1mu}#3 \rrangle}
\newcommand{\tccstranp}[2]{{\llanglebf} #1\mspace{-1mu}\blacktriangleright\mspace{-1mu}#2 {\rranglebf}}

\newcommand{\tccsType}[1]{\mathsf{#1}}
\newcommand{\tccsTypeT}{\tccsType{t}}
\newcommand{\tccsTypeP}{\tccsType{p}}
\newcommand{\tccsTypeC}{\tccsType{c}}

\newcommand{\encodingT}{\eta}
\newcommand{\encodingA}{\varsigma}
\newcommand{\encodingC}{\varrho}

\hypersetup{
	colorlinks=true,
	linkcolor=black,
	citecolor=black,
	filecolor=black,
	urlcolor=black,
	pdftitle={Distributed execution of bigraphical reactive systems},
	pdfauthor={Marino Miculan and Marco  Peressotti and Andrea Toneguzzo},
	pdfsubject={Open transactions on shared memory},
	pdfkeywords={}
}

\title{Open Transactions on Shared Memory}

\author{
	\begin{tabular}{ccccc}
	Marino Miculan&\quad& Marco  Peressotti &\quad& Andrea Toneguzzo\\
	\small\href{mailto:marino.miculan@uniud.it}{\tt marino.miculan@uniud.it}
	&&
	\small\href{mailto:marco.peressotti@uniud.it}{\tt marco.peressotti@uniud.it}
	&&
	\small\href{mailto:toneguzzo.andrea@spes.uniud.it}{\tt toneguzzo.andrea@spes.uniud.it}
	\end{tabular}\\[8pt]
	\small	Laboratory of Models and Applications of Distributed Systems \\[-.8ex]
	\small	Department of Mathematics and Computer Science\\[-.8ex]
	\small	University of Udine, Italy\\
}
\date{}

\begin{document}
\maketitle

\begin{abstract}
  \emph{Transactional memory} has arisen as a good way for solving
  many of the issues of lock-based programming.  However, most
  implementations admit \emph{isolated} transactions only, which are
  not adequate when we have to coordinate \emph{communicating}
  processes.  To this end, in this paper we present \octm, an
  Haskell-like language with \emph{open} transactions over shared
  transactional memory: processes can \emph{join} transactions at
  runtime just by accessing to shared variables.  Thus a transaction
  can co-operate with the environment through shared variables, but if
  it is rolled-back, also all its effects on the environment are
  retracted.  For proving the expressive power of \octm we give an
  implementation of \tccsm, a CCS-like calculus with open
  transactions.
\end{abstract}

\section{Introduction}
Coordination of concurrent programs is notoriously difficult.
Traditional fine-grained lock-based mechanisms are deadlock-prone,
inefficient, not composable and not scalable.  For these reasons,
\emph{Software Transactional Memory} (STM) has been proposed as a more
effective abstraction for concurrent programming
\cite{st:dc1997,ahi:popl2008,hs:sigact2012}.  The idea is to mark
blocks of code as ``atomic''; at runtime, these blocks are executed
so that the well-known ACID properties are guaranteed.  Transactions
ensure deadlock freedom, no priority inversion, automatic roll-back on
exceptions or timeouts, and greater parallelizability.  Among other
implementations, we mention \emph{STM Haskell} \cite{hmpm:ppopp2005},
which allows atomic blocks to be composed into larger ones.  STM
Haskell adopts an \emph{optimistic} evaluation strategy: the blocks
are allowed to run concurrently, and eventually if an interference is
detected a transaction is \emph{aborted} and its effects on the memory
are rolled back.

However, standard ACID transactions are still inadequate when we have
to deal with \emph{communicating} processes, i.e., which can exchange
information \emph{during} the transactions.  This is very common in
concurrent distributed programming, like in service-oriented
architectures, where processes dynamically combine to form a
transaction, and all have to either commit or abort together.  In this
scenario the participants cannot be enclosed in one transaction
beforehand, because transactions are formed at runtime.  To circumvent
this issue, various forms of \emph{open transactions} have been
proposed, where the Isolation requirement is relaxed
\cite{dk:rccs,bmm:nestedcommits,vkh:concur2010,lp:ppopp2011,ksh:fossacs2014}.
In particular, \emph{TransCCS} and \tccsm are
two CCS-like calculi recently introduced to model communicating
transactions \cite{vkh:concur2010,vkh:aplas2010,ksh:fossacs2014}.
These calculi offer methodologies for proving important properties,
such as fair-testing for proving liveness and bisimulations for
proving contextual equivalences.

Now, if we try to implement cross-transaction communications \emph{a
  la} \tccsm in STM Haskell or similar languages, it turns out that
isolated transactions are not expressive enough. As an example, let us
consider two \tccsm transactions
$\tccstranp{\bar{c}.P}{0}\mid\tccstranp{c.Q}{0}$ synchronizing on a
channel $c$.  Following the standard practice, we could implement this
synchronization as two parallel processes using a pair of semaphores
\texttt{c1,c2} (which are easily realized in STM Haskell):
\\[1.5ex]
\hfill\begin{tabular}{l|l}
$\tccstranp{\bar{c}.P}{0}=$
\begin{minipage}[t]{0.35\textwidth}
\begin{verbatim}
atomic { 
  up c1     -- 1.1
  down c2   -- 1.2
  P
}
\end{verbatim}
\end{minipage}
&\quad
 $\tccstranp{c.Q}{0}=$
\begin{minipage}[t]{0.3\textwidth}
\begin{verbatim}
atomic { 
  down c1   -- 2.1
  up c2     -- 2.2
  Q
}
\end{verbatim}
\end{minipage}
\end{tabular}\hfill
\\[1.5ex]
This implementation is going to deadlock: the only possible execution
order is 1.1-2.1-2.2-1.2, which is possible outside transactions but
it is forbidden for ACID transactions\footnote{This possibility was
  pointed out also in \cite{hmpm:ppopp2005}: ``two threads can easily
  deadlock if each awaits some communication from the other''.}.  The
problem is that ordinary STM transactions are kept isolated, while in
\tccsm they can merge at runtime.

In order to address this issue, in this paper we introduce software
transactional memory with \emph{open} transactions: processes can
\emph{join} transactions and transactions can \emph{merge} at runtime,
when they access to shared variables.  To this end, we present \octm,
a higher-order language extending the concurrency model of STM Haskell
with composable \emph{open (multi-thread)} transactions interacting
via \emph{shared memory}.  The key step is to separate the isolation
aspect from atomicity: in \octm the \texttt{atomic} construct
ensures ``all-or-nothing'' execution, but not isolation; when needed,
isolated execution can be guaranteed by a new constructor
\texttt{isolated}.  An \texttt{atomic} block is a \emph{participant}
(possibly the only one) of a transaction. Notice that transaction
merging is implicitly triggered by accessing to shared memory, without
any explicit operation or \emph{a priori} coordination.  For instance,
in \octm the two transactions of the example above would merge
becoming two participants of the same transaction, hence the two
threads can synchronize and proceed.  In order to prove formally the
expressivity of open memory transactions, we define an implementation
of \tccsm in \octm, which is proved to correctly preserve behaviours
by means of a suitable notion of simulation.  We have based our work
on STM Haskell as a paradigmatic example, but this approach is 
general and can be applied to other STM implementations.

Lesani and Palsberg \cite{lp:ppopp2011} have proposed transactions
communicating through transactional message-based channels called
\emph{transactional events}.  These mechanisms are closer to models
like TransCCS and \tccsm, but on the other hand they induce a
\emph{strict coupling} between processes, which sometimes is neither
advisable nor easy to implement (e.g., when we do not know all
transaction's participants beforehand).  In fact, most STM
implementations (including STM Haskell) adopt the shared memory model
of multi-thread programming; this model is also more amenable to
implementation on modern multi-core hardware architectures with
transactional memory \cite{moss:transactionalmemorybook}. For these
reasons, in \octm we have preferred to stick to \emph{loosely
  coupled} interactions based on shared memory only.

The rest of the paper is structured as follows.  In
Section~\ref{sec:ccctm} we describe the syntax and semantics of
\octm.  Some examples are in Section \ref{sec:examples}.  In Section
\ref{sec:express} we assess the expressiveness of \octm by providing
an implementation of \tccsm, our reference model for open
transactions.  Conclusions and directions for future work are in
Section~\ref{sec:concl}. \assumeapx{Longer proofs are in the
  Appendix}.

\section{\octm: Open Concurrent Transactional Memory}\label{sec:ccctm}

In this section we introduce the syntax and semantics of \octm, a
higher-order functional language with threads and open transaction on
shared memory. The syntax is Haskell-like (in the wake of existing
works on software transactional memories such as
\cite{hmpm:ppopp2005}) and the semantics is a
small-step operational semantics given by two relations:
$\xrightarrow{\beta}$ models transaction auxiliary operations
(e.g.~creation) while $\xrightarrow{}$ models actual term
evaluations. Executions proceeds by repeatedly choosing a thread and
executing a single (optionally transactional) operation; transitions
from different threads may be arbitrarily interleaved as long as
atomicity and isolation are not violated where imposed by the program.

\begin{figure}[t]
\centering
	\begin{tabular}{lrl}
		Value & $V \mathop{::=}$ & 
		$r \mid \lambda x. M 
		\mid \ctmreturn M
		\mid \ctmBind M N
		\mid $ 
		\\ & &$ \ctmnew M 
		\mid \ctmread r
		\mid \ctmwrite r M
		\mid $ 
		\\ & & $ \ctmfork M
		\mid \ctmatomic M N 
		\mid \ctmiso M
		\mid \ctmabort M 
		\mid \ctmretry$
		\\
		Term & $M, N \mathop{::=}$ & $x \mid V \mid M\kern.1ex N \mid \dots$
	\end{tabular}
	
	\caption{Syntax of \octm values and terms.}
	\label{fig:ctm-syntax}
	
\end{figure}

\subsection{Syntax}

 The syntax can be found in Figure~\ref{fig:ctm-syntax}
where the meta-variables $r$ and $x$ range over a given 
countable
set of locations $\Loc$ and variables $\Var$ respectively.
Terms and values are inspired to Haskell and are entirely
conventional\footnote{%
	We treat the application of monadic combinators
	(e.g.~$\ctmreturn\!$) as values in the line of  similar
	works \cite{hmpm:ppopp2005}.};
they include abstractions, application, monadic operators
($\ctmreturn{\!}$ and $\!\ctmbind\!$), memory operators
($\ctmnew\!$, $\ctmread\!$, $\ctmwrite\! \!\!$), forks,
transactional execution modalities ($\ctmatomic{\!}{\!}\!$ and 
$\ctmiso\!$) and transaction operators ($\ctmabort\!$ and $\ctmretry$).

Effectfull expressions such as $\ctmfork\!$ or $\ctmiso\!$ are glued
together by the (overloaded) monadic bind $\ctmbind$ e.g.:
\[
	\ctmnew{\textcode{0}} \ctmbind \lambda x.\bl\ctmfork{\bl\ctmwrite{x}{\textcode{42}}\br}
	\ctmbind \lambda y.\ctmread{x}\br
\]
whereas values are ``passed on'' by the monadic unit $\ctmreturn\!$.
	
Akin to Haskell, 
we will use underscores in place of unused variables
(e.g.~$\lambda\_.\textcode{0}$) and $M \ctmbnd N$
as a shorthand for $M \ctmbind \lambda\_.N$,
and the convenient \emph{\textcode{do}-notation}:
\begin{align*}
	\mathtt{do\{} x \from M \mathtt{;} N \mathtt{\}} 
	\equiv
	M \ctmbind (\lambda x . \mathtt{do\{} N \mathtt{\}})
	\qquad
	\mathtt{do\{}M \mathtt{;} N \mathtt{\}} 
	\equiv
	M \ctmbind (\lambda \_ . \mathtt{do\{} N \mathtt{\}})
	\qquad
	\mathtt{do\{}M\mathtt{\}} 
	\equiv
	M
\end{align*}
possibly trading semicolons and brackets for the conventional
Haskell \emph{layout}. For instance, the above example is rendered as

\begin{Verbatim}[tabsize = 3, commandchars=\\\{\}, codes={\catcode`$=3}]
	do 
		$x \from$ newVar 0
		fork (writeVar $x$ 42)
		readVar $x$
\end{Verbatim}

\subsection{Operational Semantics}
We present the operational semantics of \octm
in terms of an abstract machine whose states
are triples $\ctmSt{\Theta,\Delta}{P}$ formed by
\begin{itemize}
	\item
		thread family (process) $P$;
	\item 
		heap memory $\Theta : \Loc \rightharpoonup \Term$;
	\item 
		distributed working memory $\Delta : \Loc \rightharpoonup \Term \times \TrName$
\end{itemize}
where $\Term$ denotes the set of \octm terms
(cf.~Figure~\ref{fig:ctm-syntax}) and $\TrName$ 
denotes the set of names used by the
machine to identify active transactions.

We shall denote the set of all possible states as $\State$.

\paragraph{Threads}

Threads are the smaller unit of execution the machine scheduler
operates on; they execute OCTM terms and do not have any private
transactional memory.

Threads are given unique identifiers (ranged over by $t$ or variations
thereof) and, whenever they take part to some transaction, the
transaction identifier (ranged over $k,j$ or variations
thereof). Threads of the former case are represented by
$\ctmthread{M}$ where $M$ is the term being evaluated and the
subscript $t$ is the thread identifier.  Threads of the latter case
have two forms: $\ctmthreadtr{M}{M'}{N}{t}{k}$, called  and
$\ctmthreadtw{M}{M'}{t}{k}$ where:
\begin{itemize}
	\item $M$
		is the term being evaluated inside the transaction $k$; 
	\item $M'$
		is the term being evaluated as \emph{compensation} 
		in case $k$ is aborted;
	\item $N$
		is the term being evaluated as \emph{continuation} 
		after $k$ commits or aborts.
\end{itemize}
Threads with a continuation are called \emph{primary participants (to
  transaction $k$)}, while threads without continuation are the
\emph{secondary participants}.  The former group includes all and only
the threads that started a transaction (i.e.~those evaluated in an
$\textcode{atomic}$), while the latter group encompasses
threads forked inside a transaction and threads forced to join a
transaction (from outside a transactional context) because of memory
interactions.  While threads of both groups can force a transaction to
abort or restart, only primary participants can vote for its
commit and hence pass the transaction result to the continuation.

We shall present thread families using the evocative CCS-like
parallel operator $\parallel$ (cf.~Figure~\ref{fig:ctm-contextes})
which is commutative and associative.  Notice that this operator is
well-defined only on operands whose thread identifiers are distinct.
The notation is extended to thread families with $\mathbf{0}$ denoting
the empty family.

\paragraph{Memory}
The memory is divided in the heap $\Theta$ and in a distributed
working memory $\Delta$. As for traditional closed (acid) transactions
(e.g.~\cite{hmpm:ppopp2005}), operations inside a transaction
are evaluated against $\Delta$ and effects are propagated to $\Theta$
only on commits.  When a thread inside a transaction $k$ accesses a
location outside $\Delta$ the location is \emph{claimed for $k$} and
remains claimed for the rest of $k$ execution. Threads inside a
transaction can interact only with locations claimed by their
transaction. To this end, threads outside any transaction can join an
existing one and different active transactions can be merged to share
their claimed locations.

We shall denote the pair $\langle \Theta, \Delta \rangle$ by $\Sigma$
and reference to each projected component by a subscript e.g. 
$\Sigma_\Theta$ for the heap.
When describing  updates to the state $\Sigma$, we adopt the convention that
$\Sigma'$ has to be intended as equal to $\Sigma$ except if stated otherwise,
i.e. by statements like ${\Sigma'_\Theta = \Sigma_\Theta[r \mapsto M]}$. 

Formally, updates to location content are
defined on $\Theta$ and $\Delta$ as follows:

\begin{gather*}
	\Theta[r \mapsto M](s) \defeq
	\begin{cases}
		M & \text{if } r = s\\
		\Theta(s) & \text{otherwise}
	\end{cases}
	\qquad
	\Delta[r \mapsto (M,k)](s) \defeq
	\begin{cases}
		(M,k) & \text{if } r = s\\
		\Delta(s) & \text{otherwise}
	\end{cases}
\end{gather*}
for any $r,s \in \Loc$, $M \in \Term$ and $k \in \TrName$.
Likewise, updates on transaction names are defined on $\Sigma$ and $\Delta$ as follows:

\[
\Sigma[k\mapsto j] \defeq (\Theta, \Delta[k \mapsto j])
\qquad
(\Delta[k \mapsto j])(r) \defeq
\begin{cases}
\Delta(r) & \text{if}\ \Delta(r)= (M,l),l\neq k\\
(M,j) & \text{if}\ \Delta(r) = (M, k)
\end{cases}
\]

for any $r \in \Loc$, $M \in \Term$ and $k, j \in \TrName$.
Note that $j$ may occur in $\Delta$ resulting in the fusion
of the transactions denoted by $k$ and $j$ respectively.
Finally, $\varnothing$ denotes the empty memory
(i.e.~the completely undefined partial function).

\begin{figure}[t]
\centering
	\begin{tabular}{lrl}
		Thread & $T_t \mathop{::=}$ & $\ctmthread{M} \mid \ctmthreadtr{M}{M'}{N}{t}{k} \mid \ctmthreadtw{M}{M'}{t}{k}$
		\\
		Thread family & $P \mathop{::=}$ & $T_{t_1} \parallel \dots \parallel T_{t_n}\quad \forall i,j\ t_i \neq t_j  $
		\\
		Expressions &$\mathbb E\mathop{::=}$&
		$\ctxhole \mid \mathbb E \ctmbind M$
		\\
		Processes &$\mathbb P_t \mathop{::=}$&
		$\ctmthread{\mathbb E} $
		\\
		Transactions &$\mathbb T_{t,k} \mathop{::=}$&
		$\ctmthreadtr{\mathbb E}{M}{N}{t}{k}  \mid \ctmthreadtw{\mathbb E}{M}{t}{k}$
	\end{tabular}
	
	\caption{Threads and evaluation contexts.}
	\label{fig:ctm-contextes}
	
\end{figure} 

\begin{figure}[t]
	\centering
	\begin{gather*}
		\frac{M \not\equiv V \quad \mathcal V[M] = V}{M \evalTermArrow V}
		\rtlabel{rule:ctm-admn-eval}{Eval}
		\qquad
		\frac{}{ \ctmreturn M \ctmbind N \evalTermArrow N\kern.1ex M}
		\rtlabel{rule:ctm-admn-bind}{BindReturn}	
		\\
		\frac{}{\ctmretry \ctmbind M \evalTermArrow \ctmretry}
		\rtlabel{rule:ctm-admn-retry}{BindRetry}	
		\qquad
		\frac{}{ \ctmabort N \ctmbind M \evalTermArrow \ctmabort N}
		\rtlabel{rule:ctm-admn-abort}{BindAbort}
	\end{gather*}
	
	\caption{\octm semantics: rules for term evaluation.}
	\label{fig:ctm-semantics-eval}
	
\end{figure}

\begin{figure}[!t]
	\centering
	\begin{gather*}
		\frac{
			M \evalTermArrow N
		}{
			\ctmSt{\Sigma}{\ctxP{M} \parallel P} \trArrow 
			\ctmSt{\Sigma}{\ctxP{N} \parallel P}
		}
		\rtlabel{rule:ctm-pterm}{TermP}
		\qquad
		\frac{
			M \evalTermArrow N
		}{
			\ctmSt{\Sigma}{\ctxT{M}\parallel P} \trArrow \ctmSt{\Sigma}{\ctxT{N}\parallel P}
		}
		\rtlabel{rule:ctm-aterm}{TermT}
		\\
		\frac{
			t'\notin \threadnames{P} \quad t \neq t'
		}{
		\ctmSt{\Sigma}{\ctxP{\ctmfork M} \parallel P}
		\trArrow
		\ctmSt{\Sigma}{\ctxP{\ctmreturn t'} \parallel \ctmthread[t']{M} \parallel P}
		}
		\rtlabel{rule:ctm-pfork}{ForkP}
		\\
		\frac{
			t'\notin \threadnames{P} \quad t \neq t'
		}{
		\ctmSt{\Sigma}{\ctxT{\ctmfork M} \parallel P }
		\trArrow
		\ctmSt{\Sigma}{\ctxT{\ctmreturn t'} \parallel \ctmthreadtw{M}{\ctmreturn{\!}}{t'}{k} \parallel P}
		}
		\rtlabel{rule:ctm-afork}{ForkT}
		\\[.5ex]
		\threadnames{T_{t_1} \parallel \dots \parallel T_{t_n}}
		\defeq \{t_1,\dots,t_n\}      
		\\[1ex]
		\frac{
			r \notin \dom(\Sigma_\Theta)\cup\dom(\Sigma_\Delta)\quad
			\Sigma_\Theta' = \Sigma_\Theta[r\mapsto M]
		}{
			\ctmSt{\Sigma}{\ctxP{\ctmnew M} \parallel P }\trArrow \ctmSt{\Sigma'}{\ctxP{\ctmreturn r} \parallel P}
		}
		\rtlabel{rule:ctm-pnew}{NewP}		
		\\
		\frac{
			r \notin \dom(\Sigma_\Theta)\cup\dom(\Sigma_\Delta)\quad
			\Sigma_\Delta' = \Sigma_\Delta[r\mapsto (M,k)]
		}{
			\ctmSt{\Sigma}{\ctxT{\ctmnew M} \parallel P }
			\trArrow
			\ctmSt{\Sigma'}{\ctxT{\ctmreturn r} \parallel P}
		}
		\rtlabel{rule:ctm-anew}{NewT}		
		\\
		\frac{
			r \notin \dom(\Sigma_\Delta) \quad 
			\Sigma_\Theta(r) = M
		}{
		\ctmSt{\Sigma}{\ctxP{\ctmread r} \parallel P}\trArrow
		\ctmSt{\Sigma}{\ctxP{\ctmreturn M} \parallel P}
		}
		\rtlabel{rule:ctm-preadmiss}{ReadP}		
		\\
		\frac{
			r \notin \dom(\Sigma_\Delta) \quad 
			\Sigma_\Theta(r) = M \quad 
			\Sigma_\Delta' = \Sigma_\Delta[r \mapsto (M,k)]
		}{
		\ctmSt{\Sigma}{\ctxT{\ctmread r} \parallel P }\trArrow
		\ctmSt{\Sigma'}{\ctxT{\ctmreturn M} \parallel P}
		}
		\rtlabel{rule:ctm-areadmiss}{ReadT}
		\\
		\frac{
			M = \ctxE{ \ctmread r} \quad
			\Sigma_\Delta(r) = (M',k) 
		}{
		\ctmSt{\Sigma}{\ctmthread{M} \parallel P }
		\trArrow
		\ctmSt{\Sigma}{\ctmthreadtw{ \ctxE{\ctmreturn M'}}{\lambda_.M}{t}{k} \parallel P}
		}
		\rtlabel{rule:ctm-preadhit}{ReadJoin}
		\\
		\frac{
			\Sigma_\Delta(r) = (M,j) \quad
			\Sigma' = \Sigma[k \mapsto j]
		}{
			\ctmSt{\Sigma}{\ctxT{\ctmread r} \parallel P}
			\trArrow
			\ctmSt{\Sigma'}{\ctxT[t,j]{\ctmreturn M} \parallel P[k \mapsto j]}
		}
		\rtlabel{rule:ctm-areadhit}{ReadMerge}
		\\
		\frac{
			r \notin \dom(\Sigma_\Delta) \quad 
			\Sigma_\Theta(r) = N \quad 
			\Sigma_\Theta' = \Sigma_\Theta[r \mapsto M]
		}{
		\ctmSt{\Sigma}{\ctxP{\ctmwrite r M} \parallel P }\trArrow
		\ctmSt{\Sigma'}{\ctxP{\ctmreturn \void} \parallel P}
		}
		\rtlabel{rule:ctm-pwritemiss}{WriteP}
		\\
		\frac{
			r \notin \dom(\Sigma_\Delta) \quad 
			\Sigma_\Theta(r) = N \quad 
			\Sigma_\Delta' = \Sigma_\Delta[r \mapsto (M,k)]
		}{
		\ctmSt{\Sigma}{\ctxT{\ctmwrite r M} \parallel P }\trArrow
		\ctmSt{\Sigma'}{\ctxT{\ctmreturn \void} \parallel P}
		}
		\rtlabel{rule:ctm-awritemiss}{WriteT}
		\\
		\frac{
			M = \ctxE{\ctmwrite{r}{M'}} \quad
			\Sigma_\Delta(r) = (M'',k) \quad 
			\Sigma_\Delta' = \Sigma_\Delta[r \mapsto (M',k)] \quad 
		}{
			\ctmSt{\Sigma}{\ctmthread{M} \parallel P }\trArrow
			\ctmSt{\Sigma'}{\ctmthreadtw{\ctxE{\ctmreturn \void}}{\lambda\_.M}{t}{k} \parallel P}
		}
		\rtlabel{rule:ctm-pwritehit}{WriteJoin}
		\\
		\frac{
			\Sigma_\Delta(r) = (N,j) \quad
			\Sigma' = \Sigma[k \mapsto j] \quad
			\Sigma'_\Delta = \Sigma_\Delta[k \mapsto (M,j)]
		}{
			\ctmSt{\Sigma}{\ctxT{\ctmwrite r M} \parallel P }\trArrow
			\ctmSt{\Sigma'}{\ctxT[t,j]{\ctmreturn \void} \parallel P[k \mapsto j]}
		}
		\rtlabel{rule:ctm-awritehit}{WriteMerge}		
	\end{gather*}
	\vspace{-4ex}
	\caption{\octm semantics: rules for $\trArrow$.}
	\label{fig:ctm-semantics-exec}
\end{figure}

\begin{figure}[!t]
	\centering
	\begin{gather*}
	\frac{
		k \notin \transactionnames{P}
	}
	{
		\ctmSt{\Sigma}{\ctmthread{\ctmatomic{M}{N} \ctmbind N'} \parallel P}\newArrow
		\ctmSt{\Sigma}{\ctmthreadtr{M}{N}{N'}{t}{k} \parallel P}
	}
	\rtlabel{rule:ctm-atomic}{Atomic}
	\\
	\frac{
		\ctmSt{\Sigma}{\ctmthread{M}} \evalTermArrow^* \ctmSt{\Sigma'}{\ctmthread{\ctmreturn{N}}}
	}
	{
		\ctmSt{\Sigma}{\ctxP{\ctmiso M}} \evalTermArrow \ctmSt{\Sigma'}{\ctxP{\ctmreturn{N}}}
	}
	\rtlabel{rule:ctm-isop}{IsolatedP}
	\\
	\frac{		
		\textcode{op} \in \{\textcode{abort}, \textcode{return}\}
		\quad 
		\ctmSt{\Sigma}{\ctmthreadtw{M}{\textcode{return}}{t}{k}} 
		\evalTermArrow^*
		\ctmSt{\Sigma'}{\ctmthreadtw{\textcode{op}\; N}{\textcode{return}}{t}{k}}
	}{
		\ctmSt{\Sigma}{\ctxT{\ctmiso M}} \evalTermArrow \ctmSt{\Sigma'}{\ctxT{\textcode{op}\; N}}
	}
	\rtlabel{rule:ctm-isot}{IsolatedT}
	\\
	\frac{
		\Sigma'_\Delta = \mathsf{clean}(k,\Sigma_\Delta) \quad 
	}{
		\ctmSt{\Sigma}{\ctmthreadtr{\ctmabort M}{N}{N'}{t}{k}}\abArrow{M}
		\ctmSt{\Sigma'}{\ctmthread{N(M) \ctmbind N'}}}
	\rtlabel{rule:ctm-abort}{RaiseAbort1}
	\\
	\frac{
		\Sigma'_\Delta = \mathsf{clean}(k,\Sigma_\Delta) \quad 
	}{
	\ctmSt{\Sigma}{\ctmthreadtw{\ctmabort M}{N}{t}{k}}\abArrow{M}
	\ctmSt{\Sigma'}{\ctmthread{N(M)}}}
	\rtlabel{rule:ctm-abortw}{RaiseAbort2}
	\\
	\frac{
		\Sigma'_\Delta = \mathsf{clean}(k,\Sigma_\Delta) \quad 
	}{
	\ctmSt{\Sigma}{\ctmthreadtr{M}{N}{N'}{t}{k}}\sigabArrow{M}
	\ctmSt{\Sigma'}{\ctmthread{N(M) \ctmbind N'}}}
	\rtlabel{rule:ctm-sinabort1}{SigAbort1}
	\\
	\frac{
		\Sigma'_\Delta = \mathsf{clean}(k,\Sigma_\Delta) \quad 
	}{
	\ctmSt{\Sigma}{\ctmthreadtw{M}{N}{t}{k}}\sigabArrow{M}
	\ctmSt{\Sigma'}{\ctmthread{N(M)}}}
	\rtlabel{rule:ctm-sinabort2}{SigAbort2}
	\\
	\frac{
		\ctmSt{\Sigma}{P} \abArrow{M} \ctmSt{\Sigma'}{P'}
		\quad
		\ctmSt{\Sigma}{Q} \sigabArrow{M} \ctmSt{\Sigma'}{Q'}
	}
	{
		\ctmSt{\Sigma}{P\parallel Q} \abArrow{M} \ctmSt{\Sigma'}{P'\parallel Q'}
	}
	\rtlabel{rule:ctm-abbroadcast}{AbBroadcast}
	\\
	\frac{
		\Sigma_\Theta' = \mathsf{commit}(k,\Sigma_\Theta,\Sigma_\Delta) \quad
		\Sigma_\Delta' = \mathsf{clean}(k,\Sigma_\Delta) 
	}{
		\ctmSt{\Sigma}{\ctmthreadtr{\ctmreturn M}{N}{N'}{t}{k} }\coArrow
		\ctmSt{\Sigma'}{\ctmthread{\ctmreturn{M} \ctmbind N'}}
	}
	\rtlabel{rule:ctm-commitc}{Commit1}
	\\
	\frac{
		\Sigma_\Theta' = \mathsf{commit}(k,\Sigma_\Theta,\Sigma_\Delta) \quad
		\Sigma_\Delta' = \mathsf{clean}(k,\Sigma_\Delta) 
	}{
		\ctmSt{\Sigma}{\ctmthreadtw{M}{N}{t}{k} }\coArrow
		\ctmSt{\Sigma'}{\ctmthread{M}}
	}
	\rtlabel{rule:ctm-commitw}{Commit2}
	\\
	\frac{
		\ctmSt{\Sigma}{P}\coArrow\ctmSt{\Sigma'}{P'}\quad
		\ctmSt{\Sigma}{Q}\coArrow\ctmSt{\Sigma'}{Q'}
	}{
		\ctmSt{\Sigma}{P \parallel Q} \coArrow \ctmSt{\Sigma'}{P' \parallel Q'}	
	}
	\rtlabel{rule:ctm-trbroadcast}{CoBroadcast}
	\\
	\frac{
		\ctmSt{\Sigma}{P}\betaArrow\ctmSt{\Sigma'}{P'}\quad
		 \transactionnames{\beta} \notin  \transactionnames{Q}
	}{
		\ctmSt{\Sigma}{P \parallel Q} \betaArrow \ctmSt{\Sigma}{P' \parallel Q}	
	}
	\rtlabel{rule:ctm-trignore}{TrIgnore}
	\\[1ex]
	\mathsf{clean}(k,\Delta)(r) \defeq 
		\begin{cases}
			\perp &\!\text{if } \Delta(r) = (M,k)\\
			\Delta(r) & \!\text{otherwise}
		\end{cases}
	\qquad
	\mathsf{commit}(k,\Theta,\Delta)(r) \defeq 
		\begin{cases}
			M &\!\text{if } \Delta(r) = (M,k)\\
			\Theta(r) & \!\text{otherwise}
		\end{cases}
	\\[.5ex]
	\transactionnames{\ctmthread{M}} \defeq \emptyset
	\quad
	\transactionnames{\ctmthreadtr{M}{M'}{N}{t}{k}} \defeq \{k\}
	\quad
	\transactionnames{\ctmthreadtw{M}{N}{t}{k}} \defeq \{k\}
	\\
	\transactionnames{T_{t_1} \parallel \dots \parallel T_{t_n}}
		\defeq \bigcup_{1 \leq i \leq n} \transactionnames{T_{t_i}}
	\end{gather*}
	\vspace{-4ex}
	\caption{\octm semantics: rules for $\betaArrow$.}
	\label{fig:ctm-semantics-aux}
\end{figure}

\paragraph{Behaviour}
Evaluation contexts are shown in Figure~\ref{fig:ctm-contextes}
and the transition relations are presented
in Figures~\ref{fig:ctm-semantics-eval}, \ref{fig:ctm-semantics-exec},
\ref{fig:ctm-semantics-aux}.  
The first (cf.~Figures~\ref{fig:ctm-semantics-eval}) is defined on terms only
and models pure computations.

In particular, rule \eqref{rule:ctm-admn-eval} allows
a term $M$ that is not a value to be evaluated by an auxiliary
(partial) function, $\mathcal{V}[M]$ yielding the value $V$ of $M$
whereas the other three rules define the semantic of the monadic
bind.
The transition relation modelling pure computations can be thought
as accessory to the remaining two for these model transitions
between the states of the machine under definition.

Derivation rules in
Figure~\ref{fig:ctm-semantics-exec} characterize the execution of
pure (effect-free) terms, forks and memory operations both inside, and
outside of some transaction; Derivation rules in
Figure~\ref{fig:ctm-semantics-aux} characterize auxiliary operations
for transaction management (e.g.~creation) and their coordination
(e.g~distributed commits).  Note that there are no derivation rules
for $\textcode{retry}$. In fact, the meaning of $\textcode{retry}$ is
to inform the machine that choices made by the scheduler led to a
state from which the program cannot proceed.  From an implementation
perspective this translates in the transaction being re-executed from the
beginning (or a suitable check-point) following a different scheduling
of its operations.

We shall describe now a representative subset of the derivation rules
from Figures~\ref{fig:ctm-semantics-exec} and \ref{fig:ctm-semantics-aux}.

Reading a location falls into four cases depending on the location
being claimed (i.e.~occurring in $\Delta$) and the reader being part
of a transaction.  The rule \eqref{rule:ctm-preadmiss} characterize
the reading of an unclaimed location from outside any transaction; the
read is performed as expected leaving it unclaimed.  Rule
\eqref{rule:ctm-areadmiss} describes the reading of an unclaimed
location $r$ by a thread belonging to some transaction $k$; the side
effect of the reading is $r$ being claimed for $k$.  Rules
\eqref{rule:ctm-areadhit} and \eqref{rule:ctm-preadhit} cover the
cases of readings against claimed locations. In the first scenario,
the reading thread belongs to a transaction resulting in the two being
merged, which is expressed by renaming its transaction via a
substitution. In the remaining scenario, the reading thread does not
belong to any transaction and hence joins the transaction $k$ which
claimed the location. The newly created participant does not have any
continuation since the whole term is set to be executed inside $k$;
any other choice for splitting the term singling out a compensation
would impose an artificial synchronization with the transaction
commit. For a counter example, consider executing only the read
operation inside the transaction and delaying everything after the
commit; then concurrency will be clearly reduced. Because of the same
reasoning, the whole term $M$ is taken as the compensation of the
participant.

Transactions are created by rule \eqref{rule:ctm-atomic};
threads participating in a transaction are non-deterministically 
interleaved with other threads. The stronger requirement of isolation
is offered by rules \eqref{rule:ctm-isop} and \eqref{rule:ctm-isot},
whose premises forbid thread or transaction creation.

Committing or aborting a transaction require a synchronization of its
participants. In particular, an abort can be read as a participant
vetoing the outcome of the transaction; this corresponds to
\eqref{rule:ctm-abort} and \eqref{rule:ctm-abortw}. The information is
then propagated by \eqref{rule:ctm-abbroadcast} and
\eqref{rule:ctm-trignore} to any other participant to the transaction
being aborted; these participants abort performing a transition
described by either \eqref{rule:ctm-sinabort1} or
\eqref{rule:ctm-sinabort2}.

\section{Examples}\label{sec:examples}
In this section we provide some short examples to illustrate the use
of $\octm$ and how standard STM behaviour can be recovered in
$\octm$ thanks to the $\ctmiso{\!}$ construct. In Section~\ref{sec:trans} we will
give an extended example by providing a translation of \tccsm into
\octm.

\subsection{MVars}
One of the basic constructs offered by Concurrent Haskell are
\emph{MVars} \cite{pgf:popl1996} i.e.~mutable locations that are
either empty or contain a value of the given type parameter.
Interaction with these structures is based on two fundamental
operations: \textcode{putMVar} which fills an \textcode{MVar} if it is
empty and blocks otherwise, and \textcode{takeMVar} which empties an
\textcode{MVar} if it is full and blocks otherwise.
In \cite{hmpm:ppopp2005} \textcode{MVar}s are implemented on top of
\textcode{TVar}s (i.e.~STM Haskell transactional locations).

Following \cite{hmpm:ppopp2005} an \textcode{MVar} of type \textcode{a} is implemented on top of a \textcode{OTVar}
(our transactional locations i.e.~any $r \in \Loc$) 
holding a value of type \textcode{Maybe a}; this is a type that is either an empty value (\textcode{Nothing}) or actually holds a value of type \textcode{a} (e.g.~\textcode{Just 42}). Thus, the definition of the type \textcode{MVar a} is the following:
\begin{Verbatim}[tabsize = 2, commandchars=\\\{\}, codes={\catcode`$=3}]
	type MVar a = OTVar (Maybe a)
\end{Verbatim}
and its two constructors for creating an empty and a full location are:
\begin{Verbatim}[tabsize = 2, commandchars=\\\{\}, codes={\catcode`$=3}]
	newEmptyMVar = newVar Nothing
	newMVar $x$ = newVar (Just $x$)
\end{Verbatim}
The definition of the two basic operations is precisely the same
appearing in \cite{hmpm:ppopp2005} except for the added \ctmiso{\!}
construct for enforcing isolation.
\\[1ex]
\begin{minipage}[t]{0.4\textwidth}
\begin{Verbatim}[tabsize = 2, commandchars=\\\{\}, codes={\catcode`$=3}]
	takeMVar $v$ = isolated do
		$v \from \ctmread{v}$
		case $v$ of
			Nothing $\to$ \ctmretry
			Just $x$ $\to$ do
				writeVar $x$ Nothing
				return $x$
\end{Verbatim}
\end{minipage}
\hfill
\begin{minipage}[t]{0.45\textwidth}
\begin{Verbatim}[tabsize = 2, commandchars=\\\{\}, codes={\catcode`$=3}]
 putMVar $v$ $y$ = isolated do
		$v \from \ctmread{v}$
		case $v$ of
			Nothing $\to$ writeVar $y$ Nothing
			Just $\_$ $\to$ \ctmretry
\end{Verbatim}
\end{minipage}

\subsection{Transactional RPC}
\textcode{MVar}s can be used as simple directional channels 
with \textcode{takeMVar} and \textcode{putMVar} as receive and send. Then a bidirectional channel for a \emph{remote procedure call} is easily implemented using a pair of \textcode{MVar}s
\begin{Verbatim}[tabsize = 2, commandchars=\\\{\}, codes={\catcode`$=3}]
	type RPC a b = (MVar (CorId, a), MVar (CorId, b))
\end{Verbatim}
where \textcode{a} and \textcode{b} are the types of 
the request and response exchanged and \textcode{CorId} is a suitable type providing a correlation identifier for relating a request to its response.

Before we introduce the skeleton and stub let us define a conditional variation of the \textcode{takeMVar} accepting a boolean predicate $p$ and such that it empties the given \textcode{MVar} $v$ only if the contained value satisfies $p$
and blocks (issue a \textcode{retry}) otherwise.
\begin{Verbatim}[tabsize = 2, commandchars=\\\{\}, codes={\catcode`$=3}]
	takeMVarIf $p$ $v$ = isolated do
		$v \from \ctmread{v}$
		case $v$ of
			Nothing $\to$ \ctmretry
			Just $x$ $\to$ do
				if $p x$ then 
					writeVar $x$ Nothing \ctmbnd return $x$
				else
					retry
\end{Verbatim}
The conditional version of \textcode{takeMVar} allows us to take a response only if we know its correlation identifier and hence the call is simply:
\begin{Verbatim}[tabsize = 2, commandchars=\\\{\}, codes={\catcode`$=3}]
	rpcCall ($req$, $res$) $data$ = do
		$c \from $ newCorrelationId
		putMVar $req$ ($c$, $data$)
		$r \from $ takeMVarIf ($c$ == fst) $res$
		return (snd $r$)
\end{Verbatim}
where \textcode{fst} and \textcode{snd} are the first and second projections respectively.
Symmetrically, to provide the rpc we just need to take a
request from the \textcode{MVar} $req$ and put its response
in $res$ using the same correlation identifier:
\begin{Verbatim}[tabsize = 2, commandchars=\\\{\}, codes={\catcode`$=3}]
	rpcServe ($req$, $res$) $data$ = do
		$q \from $ takeMVar $req$
		$a \from$ doSomething (snd $q$)
		putMVar $res$ (fst $q$, $a$)
\end{Verbatim}
If any of the two parties happens to be partaking a transaction
the rpc results in the other joining the transaction effectively rendering the rpc transactional.

The above example is quite simplified (e.g.~requests could have been handled by a buffer, and the structure of \textcode{($req$, $req$)} should be hidden to the user) but serves the purpose of illustrating the difference between \octm and STM. 
handled by a buffer, and the structure of \textcode{($req$,
  $req$)}
should be hidden to the user) but serves the purpose of illustrating
the difference between \octm and STM.  In fact, the above
implementation allows the call to happen inside a transaction without
resulting into a lock as in the case of STM since isolation will
prevent the serving thread to join and provide a response.

\vspace{-1ex}

\section{Expressiveness of \octm}\label{sec:express}
In order to assess the expressive power of \octm, in this Section we
prove that it can be used to implement \tccsm, a formal model for open
transactions \cite{ksh:fossacs2014}.  We proceed as follow: first, in
Subsection~\ref{sec:tccsm} we recall \tccsm; then, the translation of
\tccsm processes into \octm states is defined in
Subsection~\ref{sec:trans}; this translation is proved to be correct
in Subsection~\ref{sec:adeq}.

\subsection{\tccsm: CCS with open transactions} \label{sec:tccsm}
\tccsm \cite{ksh:fossacs2014} is a CCS-like calculus with open flat
can synchronize even when belonging to different transactions, which
in turn are joined into a distributed one.  We refer to
\cite{ksh:fossacs2014} for a detailed description of \tccsm.
transactions: processes can synchronize even when belonging to
different transactions, which in turn are joined into a distributed
one.  We refer to \cite{ksh:fossacs2014} for a detailed description of
\tccsm.

The syntax of \tccsm is defined by the following grammar
\begin{equation}
	P \mathop{::=}
		\tccssum_{i=1}^{n} \alpha_i.P_i \mid 
		\tccsprod_{i=0}^{m} P_i \mid 
		P\backslash L \mid X \mid \mu X.P \mid
		\tccstranp{P_1}{P_2} \mid
		\tccstrana{P_1}{P_2} \mid
		\mathsf{co}.P
	\label{eq:tccs-grammar}
\end{equation}
\looseness=-1
where $\alpha_i ::= a \mid \bar a \mid \tau$, $a$ ranges over 
a given set of visible actions $A$, $L$ over subsets of $A$ 
and the bijection $(\overline{\cdot}) : A \to A$ maps 
every action to its
\emph{coaction} as usual. The calculus extends CCS with three
constructs which represent \emph{inactive} transactions, \emph{active}
transactions and \emph{commit} actions respectively. Transactions such
as $\tccstrana{P_1}{P_2}$ are formed by two processes with the former
being executed atomically and the latter being executed whenever the
transaction is aborted, i.e.~as a \emph{compensation}. Terms denoting
active transactions expose also a name ($k$ in the previous example)
which is used to track transaction fusions.  For instance, consider
the process denoted by $\tccstrana[j]{P_1}{P_2} \mid \tccstrana[k]{Q_1}{Q_2}$
where $P_1$ and $Q_1$ synchronize on some $a \in A$; the result of
this synchronization is the fusion of the transactions $j$ and $k$
i.e.~$\tccstrana[l]{P'_1}{P_2} \mid \tccstrana[l]{Q'_1}{Q_2}$. The fusion makes
explicit the dependency between $j$ and $k$ introduced by the
synchronization and ties them to agree on commits. In this sense,
$P'_1$ and $Q'_1$ are participants of a \emph{distributed transaction}
\cite{gl:tds2006}.

\looseness=-1
As in \cite{ksh:fossacs2014} we restrict ourselves to well-formed terms.
Intuitively, a term is well-formed if active transactions occur only at the
top-level and commit actions occur only in a transaction (active or inactive).
To this end we introduce a \emph{type system} for \tccsm, whose rules
are in Figure~\ref{fig:tccs-simple-types}.   Terms that cannot
occur inside a transaction have type $\tccsTypeT$, terms that cannot occur
outside a transaction have type $\tccsTypeC$, and terms without such 
restrictions have type $\tccsTypeP$; $\tau$ ranges over types.

\begin{figure}[t]
	\begin{gather*}
		\frac{\encodingA \vdash P : \tccsTypeP}{\encodingA \vdash P : \tau}
		\quad
		\frac{\encodingA \vdash P : \tccsTypeP}{\encodingA \vdash \mathsf{co} .P : \tccsTypeC}
		\quad
		\frac{\encodingA \vdash P : \tau}{\encodingA \vdash P\backslash L : \tau}
		\\
		\frac{}{\encodingA \vdash X : \encodingA(X)}
		\quad
		\frac{\encodingA[X : \tccsTypeP] \vdash P : \tccsTypeP}{\encodingA \vdash \mu X.P : \tccsTypeP}
		\quad
		\frac{\encodingA[X : \tccsTypeC] \vdash P : \tccsTypeC}{\encodingA \vdash \mu X.P : \tccsTypeC}
		\quad
		\frac{\forall i\, \encodingA \vdash P_i : \tau}{\encodingA \vdash \prod P_i : \tau}
		\quad
		\\
		\frac{\forall i\, \encodingA \vdash P_i : \tccsTypeP}{\encodingA \vdash \sum \alpha_i.P_i : \tccsTypeP}
		\quad
		\frac{\forall i\, \encodingA \vdash \alpha_i.P_i : \tccsTypeC}{\encodingA \vdash \sum \alpha_i.P_i : \tccsTypeC}
		\quad
		\frac{\encodingA \vdash P : \tccsTypeC \quad \encodingA \vdash Q : \tccsTypeP}
			{\encodingA \vdash \tccstrana{P}{Q} : \tccsTypeT}
		\quad
		\frac{\encodingA \vdash P : \tccsTypeC \quad \encodingA \vdash Q : \tccsTypeP}
			{\encodingA \vdash \tccstranp{P}{Q} : \tccsTypeP}		
	\end{gather*}
	\caption{Simple types for \tccsm.}
	\label{fig:tccs-simple-types}
\end{figure}

\begin{definition}[Well-formed \tccsm terms]
	\label{def:tccs-well-fomed}
	A \tccsm term $P$, described by the grammar in \eqref{eq:tccs-grammar}, is
	said to be \emph{well-formed} if, and only if, $\emptyset \vdash P : \tccsTypeT$.
	Well-formed terms form the set $\Proc$.
\end{definition}

\begin{figure}[t]
	\begin{gather*}
		\frac{}{\tccssum \alpha_i.P_i \xrightarrow{\alpha_i}_\varepsilon P_i}
		\rtlabel{rule:tccs-sum}{Sum}
		\quad
		\frac{P\xrightarrow{a}_\varepsilon P' \quad Q\xrightarrow{\bar a}_\varepsilon Q'}
			{P|Q \xrightarrow{\tau}_\varepsilon P'|Q'}
		\rtlabel{rule:tccs-sync}{Sync}
		\quad
		\frac{}{\mu X.P \xrightarrow{\tau}_\varepsilon P[{^{\mu X.P}\!/_{X}}]}
		\rtlabel{rule:tccs-rec}{Rec}
		\\
		\frac{P\xrightarrow{\alpha}_\sigma P' \quad  \mathsf{img}(\sigma)\cap\mathsf{tn}(Q) = \emptyset}
			{P|Q \xrightarrow{\alpha}_\sigma P'|Q[\sigma]}
		\rtlabel{rule:tccs-par}{ParL}
		\quad	
		\frac{
			\tau \neq \alpha_j
			}
			{
				\tccssum \alpha_i.P_i \xrightarrow{k(\alpha_j)}_{\varepsilon \mapsto k} 
			\tccstrana{P_j|\mathsf{co}}{\tccssum \alpha_i.P_i}
			}
		\rtlabel{rule:tccs-trsum}{TSum}
		\\
		\frac{P \xrightarrow{\alpha}_\varepsilon P' \quad  \tau \neq \alpha \quad  l \neq k}
			{\tccstrana[l]{P}{Q} \xrightarrow{k(\alpha)}_{l\mapsto k} \tccstrana{P'}{Q}}
		\rtlabel{rule:tccs-tract}{TAct}
		\quad	
		\frac{P\xrightarrow{k(a)}_{i \mapsto k} P' \quad  Q\xrightarrow{k(\bar a)}_{j \mapsto k} Q'}
			{P|Q \xrightarrow{k(\tau)}_{i,j \mapsto k} P'[j \mapsto k]|Q'[i \mapsto k]}
		\rtlabel{rule:tccs-trsync}{TSync}
		\\
		\frac{P \xrightarrow{\alpha}_\sigma P' \quad  \alpha \notin L}
			{P\backslash L \xrightarrow{\alpha}_\sigma P'\backslash L}
		\rtlabel{rule:tccs-restr}{Res}
		\quad
		\frac{P \xrightarrow{\tau}_\varepsilon P'}
			{\tccstrana{P}{Q} \xrightarrow{\tau}_\varepsilon \tccstrana{P'}{Q}}
		\rtlabel{rule:tccs-trtau}{TTau}
		\quad
		\frac{}{\tccstrana{P}{Q} \xrightarrow{\mathsf{ab} k} Q}
		\rtlabel{rule:tccs-trab}{TAb}	
		\\		
		\frac{P\xrightarrow{\beta} P'}{P\backslash L \xrightarrow{\beta} P'\backslash L}
		\rtlabel{rule:tccs-trrestr}{TRes}\!
		\quad
		\frac{k \text{ fresh}}{\tccstranp{P}{Q} \!\xrightarrow{\mathsf{new} k}\! \tccstrana{P}{Q}}
		\rtlabel{rule:tccs-trnew}{TNew}\!
		\quad	
		\frac{\exists i\, P_i = \mathsf{co}.P'_i}
			{\tccstrana{\tccsprod\! P_i}{Q} \xrightarrow{\mathsf{co} k} \Psi_{id}(P)}
		\rtlabel{rule:tccs-trco}{TCo}
		\\
		\frac{P \xrightarrow{\beta} P' \quad  Q\xrightarrow{\beta}Q' \quad \beta \neq \mathsf{new}k}
			{P|Q \xrightarrow{\beta} P'|Q'}
		\rtlabel{rule:tccs-trbcast}{TB1}
		\quad
		\frac{P \xrightarrow{\beta} P' \quad  \mathsf{tn}(\beta) \notin \mathsf{tn}(Q)}
				{P|Q \xrightarrow{\beta} P'|Q}
		\rtlabel{rule:tccs-trignore}{TB2}
		\\
		\begin{array}{r}
			\Psi_\sigma(P)\defeq\begin{cases}
				Q & \text{if } P \!=\! \mathsf{co}.Q\\
				\Psi_\sigma(Q)\backslash L & \text{if } P \!=\! Q\backslash L\\
				\tccssum \alpha_i.\Psi_\sigma(P_i) & \text{if } P \!=\! \tccssum \alpha_i.P_i\\
				\tccsprod \Psi_\sigma(P_i) & \text{if } P \!=\! \tccsprod P_i\\
				\mu X.\Psi_{\sigma[{^P\!\!/\!_X}]}(Q) & \text{if } P \!=\! \mu X.Q\\
				P[\sigma] & \text{otherwise}
			\end{cases}
		\end{array}\quad
		\begin{array}{rl}
			\mathsf{tn}(P)\defeq&\begin{cases}
				\{k\} & \text{if } P \!=\! \tccstrana{P}{Q}\\
				\bigcup \mathsf{tn}(P_i) & \text{if } P \!=\! \textstyle\tccsprod P_i \\
				\emptyset & \text{otherwise}
			\end{cases}
			\\
			\mathsf{tn}(\beta)\defeq&\begin{cases}
				k & \text{if } \beta \!=\! \mathsf{new} k\\
				k & \text{if } \beta \!=\! \mathsf{ab} k\\
				k & \text{if } \beta \!=\! \mathsf{co} k
			\end{cases}
		\end{array}
	\end{gather*}
	\caption{\tccsm operational semantics.}
	\label{fig:tccs-semantics}
\end{figure}

The operational semantics of well-formed \tccsm terms is given by the
SOS in Figure~\ref{fig:tccs-semantics} (see \cite{ksh:fossacs2014} for
further details). The reduction semantics is given as a binary
relation $\to$ defined by
\[
	P \to Q \defiff 
	P \xrightarrow{\tau}_\sigma Q \lor
	P \xrightarrow{\beta} Q \lor
	P \xrightarrow{k(\tau)}_\sigma Q
	\text{.}
\]
The first case is a synchronization between pure CCS processes. The
second case corresponds to creation of new transactions and
distributed commit or abort
($\beta \in \{\mathsf{new}k,\mathsf{co}k,\mathsf{ab}k\}$). The third
case corresponds to synchronizations of processes inside a named (and
possibly distributed) transaction. Notice that by
\eqref{rule:tccs-trsync} transaction fusion is \emph{driven by
  communication} and that by \eqref{rule:tccs-trsum} any pure CCS
process can join and interact with a transaction.

\subsection{Encoding \tccsm in \octm}\label{sec:trans}

In this section we define the translation from \tccsm processes to
\octm states.  To this end, we have to implement transactions and
CCS-like synchronizations using shared transactional variables and the
\texttt{atomic} and \texttt{isolated} operators.

Synchronization is implemented by means of shared transactional variables,
one for each channel, that take values of type
\textcode{ChState} (cf.~Figure~\ref{fig:channels}); 
this type has four constructors:
one for each of the three messages of the communication protocol below
plus a ``nothing'' one providing the default value.
Let $t_1$ and $t_2$ be the identifiers of two threads
simulating $a.P$ and $\overline{a}.Q$ respectively.
The protocol is composed by the following four steps:
\begin{enumerate}
	\item 
		$t_1$ checks whether the channel is free and writes on the transactional variable modelling
		the channel $a$ a nonce tagged with the constructor $\textcode{M1}$;
	\item
		$t_2$ reads the variable for $a$
		and accepts the synchronization offered by 
		the challenge \textcode{(M1 np)} adding a fresh nonce
		to it and writing back \textcode{(M2 np nq)};
	\item
		$t_1$ reads the answer to its challenge and acknowledges
		the synchronization writing back the nonce it read
		tagged with the constructor $\textcode{M3}$;
	\item
		$t_2$ reads the acknowledgement and frees the channel.
\end{enumerate}

Each step has to be executed in isolation with respect to the interactions with the shared transactional 
variable $a$.

Nonces are meant to correlate the steps only and hence can be easily
implemented in \octm by pairing thread identifiers with counter
\emph{a la} logical clock.  If at any step a thread finds the channel
in an unexpected state it means that the chosen scheduling has led to
a state incoherent with respect to the above protocol; hence the
thread executes a \emph{retry}.  This tells the scheduler to try
another execution order; by fairness, we eventually find a scheduling
such that the two processes do synchronize on $a$ and these are the
only executions leading to $P \mid Q$.
\begin{figure}[t]
	\centering
	\begin{tikzpicture}[auto,xscale=2.5,yscale=.5,
			arr/.style={shorten <=-#1},
			arr/.default=-2pt,
			msg/.style={yshift=-2pt}
		]
		
		\node (p0) at (0,6) {\(a.P\)};
		\node (q0) at (2,6) {\(\overline{a}.Q\)};
		\node (v0) at (1,6.5) {var \(a\)};
		
		\draw[gray, dashed] (p0) -- (0,-1);
		\draw[gray, dashed] (q0) -- (2,-1);
		\draw[gray, dotted] (v0) -- (1,-1);

		\draw[<-,arr] (0,5) to node[msg] {\textcode{M0}} (.9,5);
		\draw[->,arr] (0,4) to node[msg] {\textcode{(M1 $np$)}} (.9,4);
		\draw[<-,arr] (2,3.5) to node[msg,swap] {\textcode{(M1 $nx$)}} (1.1,3.5);
		
		\draw[->,arr] (2,2.5) to node[msg,swap] {\textcode{(M2 $nx$ $nq$)}} (1.1,2.5);
		\draw[<-,arr] (0,2) to node[msg] {\textcode{(M2 $np$ $ny$)}} (.9,2);
	
		\draw[->,arr] (0,1) to node[msg] {\textcode{(M3 $ny$)}} (.9,1);
		\draw[<-,arr] (2,.5) to node[msg,swap] {\textcode{(M3 $nq$)}} (1.1,.5);
		\draw[->,arr] (2,-.5) to node[msg,swap] {\textcode{M0}} (1.1,-.5);
		
		\draw [decorate,decoration={brace,amplitude=3pt},xshift=-1pt]
			(0,4) -- (0,5) node [midway,xshift=-2pt] {1};
		\draw [decorate,decoration={brace,amplitude=3pt},xshift=1pt]
			(2,3.5) -- (2,2.5) node [midway,xshift=2pt] {2};
		\draw [decorate,decoration={brace,amplitude=3pt},xshift=-1pt]
			(0,1) -- (0,2) node [midway,xshift=-2pt] {3};
		\draw [decorate,decoration={brace,amplitude=3pt},xshift=1pt]
			(2,.5) -- (2,-.5) node [midway,xshift=2pt] {4};		
	\end{tikzpicture}
	
	\caption{Implementing \tccsm synchronization.}
	\label{fig:sync-example}
	
\end{figure}
The protocol is illustrated in Figure~\ref{fig:sync-example}.
If the synchronizing parties are involved in distinct transactions these are fused as a side effect of
the interaction via the shared variable.

A choice like $\tccssum_{i = 1}^{m} \alpha_i.P_i$ can be seen as a
race of threads $t_1,\dots,t_m$, each simulating a branch, to acquire
a boolean transactional variable $l$ (private to the group).  Each
$t_i$ proceeds as follows. First, it checks $l$ and if it is set, it
returns void and terminates (another thread has already acquired it);
otherwise it tries to set it while carrying out $\alpha_i$, i.e.~right
before executing its last step of the communication protocol.  If the
variable is acquired by another thread while $t_i$ is finalizing
$\alpha_i$ then $t_i$ issues a \texttt{retry} to retract any effect of
$\alpha_i$.  The \octm code implementing this protocol is shown in
Figure~\ref{fig:channels}.

\begin{figure}[!t]
\begin{Verbatim}[tabsize = 2, commandchars=\\\{\}, codes={\catcode`$=3}]
data Channel = OTVar ChState
data ChState = M1 Nonce | M2 Nonce Nonce | M3 Nonce | M0

tau $l$ $P$ = \ctmiso do
	case ($\ctmread{l}$) of
		False $\to$ $\ctmreturn{\void}$
		True $\to$ chooseThis $l$ \ctmbnd $P$

chooseThis $l$ = \ctmwrite{$l$}{False}
			
eqOrRetry $x$ $y$ 
	| $x$ == $y$ = return \void
	| otherwise = \ctmretry

bang $x$ = fork $x$ \ctmbnd bang $x$

recv $c$ $l$ $P$ = do
	$nq$ $\from$ newNonce
	\ctmiso do
		case (\ctmread{l}) of
			False $\to$ $\ctmreturn{\void}$
			True $\to$ do
				chooseThis $l$
				case (\ctmread{$c$}) of
					(M1 $nx$) $\to$ \ctmwrite{$c$}{(M2 $nx$ $nq$)}
					_ $\to$ \ctmretry
	\ctmiso do
		case (\ctmread{$c$}) of
			(M3 $ny$) $\to$ eqOrRetry $ny$ $nq$ \ctmbnd \ctmwrite{$c$}{M0} \ctmbnd $P$
			_ $\to$ \ctmretry
			
send $c$ $l$ $P$ = do
	$np$ $\from$ newNonce
	\ctmiso do
		case (\ctmread{$l$}) of
			False $\to$ $\ctmreturn{\void}$
			True $\to$ do
				chooseThis $l$
				case (\ctmread{$c$}) of
					M0 $\to$ $\ctmwrite{c}{\bl\textcode{M1} np\br}$
					_ $\to$ \ctmretry
	\ctmiso do
		case (\ctmread{$c$}) of
			(M2 $nx$ $ny$) $\to$ eqOrRetry $nx$ $np$ \ctmbnd \ctmwrite{$c$}{(M3 $ny$)} \ctmbnd $P$
			_ $\to$ \ctmretry
\end{Verbatim}
	\caption{Encoding channels and communication}
	\label{fig:channels}
\end{figure}

\begin{figure}[!t]
\begin{minipage}[t]{0.45\textwidth}
\begin{align*}
\encodingC(\tccssum_{i = 1}^{m} \alpha_i P_i) &\defeq
\texttt{do}\\
&\quad\quad l \from \ctmnew{\textcode{True}} \\
&\quad\quad \forall i \in \{1,\dots, m\}\\
&\quad\quad\quad \ctmfork{ \xi (\alpha_i, \texttt{l}, P_i) }\\
\encodingC(\tccsprod_{i=0}^{m} P_i) &\defeq
\texttt{do}\\
&\quad\quad \forall i \in \{0,\dots, m\}\\
&\quad\quad\quad \ctmfork{\encodingC(P_i)}
\\
\encodingC(P \setminus L) &\defeq
\texttt{do}\\
&\quad\quad \forall c \in L\\
&\quad\quad\quad c \from \ctmnew{\texttt{M0}}\\
&\quad\quad \encodingC(P)
\\
\encodingC(X) &\defeq X
\\
&\quad\quad \encodingC(P)
\\
\encodingC(\mathsf{co}.P) &\defeq \texttt{do}\\
&\quad\quad l \from \ctmnew{\textcode{True}} \\
&\quad\quad\texttt{send}\ co\ l\ \encodingC(P)
\end{align*}
\end{minipage}\quad
\begin{minipage}[t]{0.45\textwidth}
\begin{align*}
\encodingC(\mu X.P) &\defeq
\texttt{let}
\ X = \encodingC(P)
\ \texttt{in}\\
\encodingC(\tccstranp{P}{Q}) &\defeq \textcode{do} \\
&\quad\quad co \from \ctmnew{\textcode{M0}}\\
&\quad\quad \ctmatomic{\texttt{p}}{\encodingC(Q)}\\
&\quad\quad \textcode{bang psi}\\
&\quad \texttt{where}\\
&\quad\quad \texttt{p} = \textcode{do}\\
&\quad\quad\quad \encodingC(P)\\
&\quad\quad\quad \ctmfork{\bl\ctmabort{\void}\br}\\
&\quad\quad\quad \textcode{psi}
\\
&\quad\quad \texttt{psi} = \textcode{do}\\
&\quad\quad\quad l \from \ctmnew{\textcode{True}}\\
&\quad\quad\quad \textcode{recv}\ co\ 
l\ \textcode{return}
\\ \xi (\alpha_i, l, P_i) &\defeq 
\begin{cases}
\textcode{recv}\ \alpha_i\ l\ \encodingC(P_i) & \text{if } \alpha_i = c \\
\textcode{send}\ \overline{\alpha_i}\ l\ \encodingC(P_i) & \text{if }\alpha_i = \overline{c} \\
\textcode{tau}\ l\ \encodingC(P_i) & \text{if }\alpha_i = \tau \\
\end{cases}
\end{align*}
\end{minipage}
	\caption{Encoding \tccsm terms of type $\tccsTypeC$}
	\label{fig:encodingC}
\end{figure}

\paragraph{Encoding of \tccsm}
We can now define the encoding $\encodingT:\Proc\to\State$, mapping
well-formed \tccsm terms to states of the \octm abstract machine.
Intuitively, a process $P \equiv \tccsprod_{i=1}^{m} P_i$ is mapped
into a state with a thread for each $P_i$ and a variable for each
channel in $P$.  Clearly a state of this form can be generated by a
single \octm term which allocates all variables and forks the $m$
threads; we have preferred to map \tccsm terms to \octm states
instead of \octm term for sake of simplicity.

The map $\encodingT$ is defined by recursion along the derivation of
$\emptyset \vdash P : \tccsTypeT$ and the number $m$ of parallel
components in $P \equiv \tccsprod_{i=1}^{m} P_i$.  This is handled by
the auxiliary encoding
$\encodingA:\Proc\times \mathsf{Heap} \to\State$ (up to choice of fresh names) whose second argument
is used to track memory allocations.
The base case is given by $m = 0$ and yields a state with no threads
i.e.~$\langle \mathbf{0}, \Theta, \varnothing\rangle$.
The recursive step is divided in three subcases depending on the structure
and type of $P_1$ ($m > 0$). 
\begin{enumerate}[]
	\item
		If ${\emptyset \vdash P_1 : \tccsTypeC}$
		without top-level restrictions (i.e.~for no $Q$ and no $L = \{a_1,\dots,a_{n+1}\}$ such that each $a_i$ occurs in $Q$ the process $P_1$ is structurally equivalent to $Q\setminus L$) then
		\[
			\encodingA(\tccsprod_{i=1}^{m+1} P_i, \Theta) \defeq
			\langle \ctmthread[t_1]{\encodingC(P_1)} \parallel S; \Sigma \rangle
		\]
		where ${\langle S;\Sigma\rangle = \encodingA(\tccsprod_{j=1}^{m-1} P_{j+1}, \Theta)}$ is the translation of the rest of $P$ and $t_1$ is unique w.r.t.~$S$ (i.e.~${t_1 \notin \threadnames{S}}$). 
		By hypothesis $P_1$ does not contain any
		top-level active transaction or parallel composition and hence 
		can be translated directly into a \octm-term 
		 
		by means of
		the encoding $\encodingC$ (cf.~Figure~\ref{fig:encodingC}) --
		$\encodingC(P)$ contain a free variable for each 
		unrestricted channel occurring in $P$.
		
	\item
		If $P_1$ has a top-level restriction
		(i.e.~${P_1 \equiv Q\setminus \{a_1,\dots,a_{n+1}\}}$) then
		\[
			\encodingA(\tccsprod_{i=1}^{m+1} P_i, \Theta) \defeq
			\langle S_1[r_1/a_1, \dots r_{n+1}/a_{n+1}] \parallel S_2; \Theta_2[r_1,\dots,r_{n+1} \mapsto \textcode{M0}],\varnothing \rangle
		\]
		where ${\langle S_1;\Theta_1,\varnothing\rangle = \encodingA(Q, \Theta)}$ and
		${\langle S_2;\Theta_2,\varnothing\rangle = \encodingA(\tccsprod_{j=1}^{m-1} P_{j+1}, \Theta_1)}$
		are the translation of the unrestricted process $Q$
		and	the translation of the rest of $P$ respectively,
		all threads have a unique identifier ${\threadnames{S_1} \cap \threadnames{S_2} = \emptyset}$,
		the heap is extended with $n$ channel variables
		fresh ($r_1,\dots,r_{n+1} \notin \dom(\Theta_2)$) and known only to
		the translation of $Q$.
	\item
		If ${P_1 \equiv \tccstrana{Q_1}{Q_2}}$ is an active transaction then
		\begin{gather*}
			\encodingA(\tccsprod_{i=1}^{m+1} P_i, \Theta) \defeq
			\langle S_{co} \parallel S_{ab} \parallel S_1[r_{co}/co] \parallel S_2; \Theta_2[r_{l} \mapsto True,r_{co} \mapsto \textcode{M0}],\varnothing \rangle
			\\
			S_{co} = \ctmthreadtr{\textcode{recv $r_l$ $r_{co}$}}{\encodingC(Q_1)}{\textcode{bang (recv (newVar True) $r_{co}$)}}{t_{co}}{k}
			\\
			S_{ab} = \ctmthreadtw{\ctmabort{\void}}{\textcode{return}}{t_{ab}}{k}
		\end{gather*}
		where 
		${\langle S_1;\Theta_1,\varnothing\rangle = \encodingA(Q_1, \Theta)}$,
		${\langle S_2;\Theta_2,\varnothing\rangle = \encodingA(\tccsprod_{j=1}^{m-1} P_{j+1}, \Theta_2)}$ (like above), the thread $S_{ab}$ is always ready to
		abort $k$ as in \eqref{rule:tccs-trab} and $S_{co}$ awaits on the
		private channel $r_{co}$ a thread from $S_1$ to reach a commit and, after
		its commit, collects all remaining synchronizations on $r_{co}$
		to emulate the effect of $\Psi$ (cf.~Figure~\ref{fig:tccs-semantics}).
		Finally, all threads have to be uniquely identified:
		$\threadnames{S_1} \cap \threadnames{S_2} = \emptyset$ and
		$t_{co},t_{ab} \notin \threadnames{S_1} \cup \threadnames{S_2}$
\end{enumerate}
\begin{remark}
  The third case of the definition above can be made more precise (at
  the cost of a longer definition) since the number of commits to be
  collected can be inferred from $Q$ mimicking the definition of
  $\Psi$.  This solution reduces the presence of dangling auxiliary
  processes and transaction fusions introduced by the cleaning
  process.
\end{remark}

Like $\encodingC$, $\encodingA(P,\Theta)$ 
contains a free variable for each unrestricted channel in $P$.
Finally, the encoding $\encodingT$ is defined on each $P \in \Proc$ as:
\[
	\encodingT(P) \defeq 
	\ctmSt{\Theta[r_1,\dots,r_{n} \mapsto \textcode{M0}], \varnothing}{S[r_1/a_1, \dots r_{n}/a_{n}]}	
\]
where $\ctmSt{\Theta,\varnothing}{S} = \encodingA(P,\varnothing)$, 
$\{r_1,\dots,r_n\} \subseteq \Loc$, and
$\{a_1,\dots,a_n\} \subseteq A$ is the set of 
channels occurring in $P$.

\subsection{Adequacy of translation}\label{sec:adeq}
In this section we prove that the translation $\encodingT$ is
adequate, in the sense that it preserves the observational behaviour
of \tccsm processes.  More precisely, akin to \cite{leroy:jar09}, we
define an appropriate notion of \emph{star simulation} $\mathcal{S}$
between well-formed \tccsm processes and states of \octm.  The basic
idea is that a single step of $P$ is simulated by a sequence of
reductions of $\encodingT(P)$, and $\encodingT(P)$ does not exhibit
behaviours which are not exhibited by $P$.
\begin{definition}[Star simulation]
\label{def:starsim}
A relation $\mathcal{S} \subseteq \Proc \times \State$ is a \emph{star
  simulation} if for all $(P,\ctmSt{\Sigma}{S})\in\mathcal{S}$:
  \begin{enumerate}
  \item for all $Q$ such that $P\xrightarrow{\tau}_\sigma Q$ or
    $P\xrightarrow{k(\tau)}_\sigma Q$, there exist $S',\Sigma'$ such
    that $\ctmSt{\Sigma}{S} \rightarrow^* \ctmSt{\Sigma'}{S'}$ and
    $(Q,\ctmSt{\Sigma'}{S'})\in\mathcal{S}$;
  \item for all $Q$ such that $P\xrightarrow{\beta} Q$, there
    exist $S',\Sigma'$ 
    s.t.~$\ctmSt{\Sigma}{S} \xrightarrow{\smash{\beta}}^* \ctmSt{\Sigma'}{S'}$ and
    $(Q,\ctmSt{\Sigma'}{S'})\in\mathcal{S}$.
  \item for all $S',\Sigma'$ such that
  $\ctmSt{\Sigma}{S}\rightarrow \ctmSt{\Sigma'}{S'}$, there exist
  $Q,S'',\Sigma''$ such that $(Q,\ctmSt{\Sigma''}{S''})\in\mathcal{S}$
  and one of the following holds:
  \begin{itemize}
  \item $P \xrightarrow{\tau}_\sigma Q$ or
    $P \xrightarrow{k(\tau)}_\sigma Q$, and
    $\ctmSt{\Sigma'}{S'} \rightarrow^* \ctmSt{\Sigma''}{S''}$

  \item $P \xrightarrow{\beta}_\epsilon Q$ and
    $\ctmSt{\Sigma'}{S'} \xrightarrow{\smash{\beta}}^*  \ctmSt{\Sigma''}{S''}$.
  \end{itemize}
\end{enumerate}
  where $\beta$-labels of the two transition relations are considered equivalent
  whenever are both commits or both aborts for the same transaction name.
We say that \emph{$P$ is star-simulated by $\ctmSt{\Sigma}{S}$} if
there exists a star-simulation $\mathcal{S}$ such that
$(P,\ctmSt{\Sigma}{S})\in\mathcal{S}$.
We denote by $\starsim$ the largest star simulation.
\end{definition}

Another technical issue is that two equivalent \tccsm processes can
be translated to \octm states which differ only on non-observable
aspects, like name renamings, terminated threads, etc.  To this end,
we need to consider \octm states up-to an equivalence relation
$\tequiv \subseteq \State \times \State$, which we define next.
\begin{definition} Two \octm states are
  \emph{transaction-equivalent}, written
  $\ctmSt{\Sigma_{1}}{S_{1}} \tequiv \ctmSt{\Sigma_{2}}{S_{2}}$, when
  they are equal up to:
\begin{itemize}
	\item renaming of transaction and thread names;
	\item \emph{terminated} threads, i.e.~threads of one of the
          following forms: $\ctmthread{\ctmreturn{M}}$,
          $\ctmthread{\ctmabort{M}}$,
          $\ctmthreadtw{\ctmreturn{\!}}{\ctmreturn{\!}}{t}{k}$,
          $\ctmthreadtw{\ctmabort{\!}}{\ctmreturn{\!}}{t}{k}$,
          $\ctmthread[t]{\textcode{psi}}$;
	\item threads blocked in synchronizations on $co$ variables.
\end{itemize}
\end{definition}
\begin{definition}
  Let $P\in Proc$ be a well-formed process and $\ctmSt{\Sigma}{S}$ be
  a state. $P$ is \emph{star simulated by $\ctmSt{\Sigma}{S}$ up to
    $\tequiv$} if $(P,\ctmSt{\Sigma}{S}) \in \starsim \circ \tequiv$.
\end{definition}

We are now ready to state our main adequacy result, which is a direct
consequence of the two next technical lemmata. 

\begin{lemma}\label{prop:1}
  For all $P,Q\in Proc$ the following hold true:
  
  \begin{enumerate}
  \item if $P\xrightarrow{\tau}_\sigma Q$ or
    $P\xrightarrow{k(\tau)}_\sigma Q$, there exist $S,\Sigma$ such
    that $\encodingT(P) \rightarrow^* \ctmSt{\Sigma}{S}$ and
    $\ctmSt{\Sigma}{S} \tequiv \encodingT(Q)$;
  \item if $P\xrightarrow{\beta} Q$, there
    exist $S,\Sigma$ such that
    $\encodingT(P) \xrightarrow{\smash{\beta}}^* \ctmSt{\Sigma}{S}$ and
    $\ctmSt{\Sigma}{S} \tequiv \encodingT(Q)$.
	\end{enumerate}	 
\end{lemma}
\begin{proofatend}
	The proof proceeds by induction on the syntax of \tccsm. 
	We only show three cases:
	\begin{enumerate}
		\item a transition $P \xrightarrow{\tau}_\varepsilon Q$ resulting from a synchronization outside a transaction;
		\item a transition $P\xrightarrow{k(\tau)}_\sigma Q$ resulting from a synchronization inside a transaction;
		\item a commit transition $P\xrightarrow{\mathsf{co}k} Q$.
	\end{enumerate}
	\begin{enumerate}
		\item If $P \xrightarrow{\tau}_\varepsilon Q$ with \eqref{rule:tccs-sync} rule, $P = P_1 \mid P_2$, $Q = Q_1 \mid Q_2$, $P_1 \xrightarrow{a} Q_1$, $P_2 \xrightarrow{\overline{a}} Q_2$.\\
		$P_1 = ((\tccssum_{i}^{m_1} a_i.R'_i)\mid P_1')\setminus L_{1} \mbox{ such that } \exists i.a_{i} = a \mbox{ and } a\notin L$\\
		$P_2 = ((\tccssum_{j}^{m_2} b_j.R''_j)\mid P_2')\setminus L_{2} \mbox{ such that } \exists j.b_{j} = \overline{a} \mbox{ and } a\notin L$	
		\begin{align*}
		\encodingT(P) = &
		\ctmSt{(\Theta,\emptyset)}
		{
			\ctmthread[t_{1}]{\ctmreturn{t'_1}} \parallel 
			\ctmthread[t_{sum1}]{\ctmreturn{t_{1m_{1}}}} \parallel
			S'_{1} \parallel
			\\
			& \parallel
			\ctmthread[t_{11}]{\xi(a_1, l_{sum1}, R'_1)} \parallel
			\dots \parallel
			\ctmthread[t_{1m_{1}}]{\xi(a_{m_{1}}, l_{sum1}, R'_{m_1})} \parallel
			\\
			&\parallel
			\ctmthread[t_{2}]{\ctmreturn{t'_2}} \parallel 
			\ctmthread[t_{sum2}]{\ctmreturn{t_{2m_{2}}}} \parallel
			S'_{2} \parallel \\
			& \parallel
			\ctmthread[t_{21}]{\xi(b_1, l_{sum2}, R''_1)} \parallel
			\dots \parallel
			\ctmthread[t_{2m_{2}}]{\xi(b_{m_{2}}, l_{sum2}, R''_{m_2})}
		}
		\end{align*}
		From hypothesis, there exists a thread $ t_{1r} \in \{t_{11},\dots, t_{1m_{1}}\}$ such that $$ \ctmthread[t_1r]{\encodingC(a_{r}.R'_{r})} = \ctmthread[t_1r]{\texttt{recv }a\ l_{sum1}\ R'_{r}} $$ and exists another thread 
		$ t_{2s} \in \{t_{21},\dots, t_{2m_{2}}\}$ s.t. $$ \ctmthread[t_2s]{\encodingC(a_{s}.R''_{s})} = \ctmthread[t_2s]{\texttt{send }a\ l_{sum2}\ R''_{s}}\text.$$		
		$l_{sum1}$ and $l_{sum2}$ are locations created by threads $t_{sum1}$ and $t_{sum2}$ from the code generated by encoding of the sums.	
		\begin{align*}
		&\ctmSt{(\Theta,\emptyset)}
		{
			\dots \parallel \ctmthread[t_1r]{\texttt{recv }a\ l_{sum1}\ \encodingC(R'_{r})} \parallel
			\ctmthread[t_2s]{\texttt{send }a\ l_{sum2}\ \encodingC(R''_{s})} \parallel \dots
		}\\
		\rightarrow^*
		&
		\ctmSt{(\Theta', \emptyset)}
		{
			\ctmthread[t_{1}]{\ctmreturn{t'_1}} \parallel 
			\ctmthread[t_{sum1}]{\ctmreturn{t_{1m_{1}}}} \parallel
			S'_{1} \parallel
			\\
			& \parallel
			\ctmthread[t_{11}]{\ctmreturn{\!}} \parallel
			\dots \parallel
			\ctmthread[t_1r]{\encodingC(R'_{r})} \parallel
			\dots \parallel
			\ctmthread[t_{1m_{1}}]{\ctmreturn{\!}} \parallel
			\\
			&\parallel
			\ctmthread[t_{2}]{\ctmreturn{t'_2}} \parallel 
			\ctmthread[t_{sum2}]{\ctmreturn{t_{2m_{2}}}} \parallel
			S'_{2} \parallel \\
			& \parallel
			\ctmthread[t_{21}]{\ctmreturn{\!}} \parallel
			\dots \parallel
			\ctmthread[t_2s]{\encodingC(R''_{s})} \parallel
			\dots \parallel
			\ctmthread[t_{2m_{2}}]{\ctmreturn{\!}}
		}= \ctmSt{\Sigma}{S}\\
		& \mbox{where } \Theta'(l_{sum1}) = \texttt{False},\ \Theta'(l_{sum2}) = \texttt{False}
		\end{align*}		
		$\texttt{recv }a\ l_{sum1}\ \encodingC(R'_{r})$ and $\texttt{send }a\ l_{sum2}\ \encodingC(R''_{s})$ can in order execute the isolated blocks, and at the end reduce to continuations $\encodingC(R'_{r})$ and $\encodingC(R''_{s})$.
		
		Other threads forked by threads $t_{sum1}, t_{sum2}$ can only reduce to $\ctmreturn{\!}$ because $\Theta'(l_{sum1}) = \texttt{False}$ and  $\Theta'(l_{sum2}) = \texttt{False}$: threads $t_{1r}, t_{2s}$ modified $l$-variables 	
		through the synchronization code inside isolated blocks.
		We can observe $ \ctmSt{\Sigma}{S} \tequiv \encodingT(Q)$, in fact $Q = (R'_{i} \mid P'_{1})\setminus L_{1} \mid (R''_{j} \mid P'_{2})\setminus L_{2}$, $\encodingT(Q) = \ctmSt{\Sigma_q}{S_q}$.	
		\begin{align*}
		\encodingT(Q) =
		& \ctmSt{(\Theta'_{q}, \emptyset)}{
			\ctmthread[t_{1}]{\ctmreturn{t'_1}} \parallel
			\ctmthread[t_{i}]{\encodingC(R'_i)} \parallel
			S'_{1} \parallel 
			\\
			&\parallel \ctmthread[t_{2}]{\ctmreturn{t'_2}} \parallel
			\ctmthread[t_{j}]{\encodingC(R'_j)} \parallel
			S'_{2}	
		} =\ctmSt{\Sigma_q}{S_q}\\
		&\mbox{where } \forall ch \in L_{1} \uplus L_{2} .\ \Theta'_{q}(ch) = \texttt{M0}
		\end{align*}
	
		$\ctmSt{\Sigma_{q}}{S_{q}}$ and $\ctmSt{\Sigma}{S}$ are different only in local variables and for reduced threads, thus $\ctmSt{\Sigma}{S} \tequiv \encodingT(Q)$.
		
		\item If $P \xrightarrow{k(\tau)}_{i,j\mapsto k} Q$ with \eqref{rule:tccs-trsync} rule, $P = (P_1 \mid P_2)$, $Q = (Q_1 \mid Q_2)$, $P_1 \xrightarrow{k(a)}_{i \mapsto k} Q_1$, $P_2 \xrightarrow{k(\overline{a})}_{j \mapsto k} Q_2$.
		
		$P = \tccstrana[i]{P_{1}}{C_{1}} \mid \tccstrana[j]{P_{2}}{C_{2}} $ and $P_{1}=P_{11}\mid\dots\mid P_{1m_{1}}$ , $P_{2}=P_{21}\mid\dots\mid P_{2m_{2}}$.
		\begin{align*}
		\encodingT(P) =&
		\ctmSt{(\Theta, \Delta)}
		{
			\ctmthreadtw{P'_{1}}{\encodingC(C_{1})}{t_{1}}{i}\parallel \ctmthreadtw{P'_{2}}{\encodingC(C_{2})}{t_{2}}{i} \parallel \ctmthreadtw{\encodingC(P'_{11})}{\ctmreturn{\!}}{t_{11}}{i} \parallel \dots 
			\\
			&\dots \parallel \ctmthreadtw{\texttt{recv}\ a\ l\ P'_{1r}}{\ctmreturn{\!}}{t_{r}}{i} \parallel \dots 
			\parallel
			\ctmthreadtw{\encodingC(P'_{1m_{1}})}{\ctmreturn{\!}}{t_{1m_{1}}}{i}
			\parallel
			\dots\\
			&\dots\parallel
			\ctmthreadtw{\encodingC(P'_{21})}{\ctmreturn{\!}}{t_{21}}{j} \parallel \dots \parallel
			\ctmthreadtw{\texttt{send}\ a\ l\ P'_{2s}}{\ctmreturn{\!}}{t_{s}}{j} \dots\\
			&\dots\parallel
			\ctmthreadtw{\encodingC(P'_{2m_{2}})}{\ctmreturn{\!}}{t_{2m_{2}}}{j}
		}
		\\
		\rightarrow^*&
		\ctmSt{(\Theta, \Delta')}
		{
			\ctmthreadtw{P'_{1}}{\encodingC(C_{1})}{t_{1}}{j}\parallel \ctmthreadtw{P'_{2}}{\encodingC(C_{2})}{t_{2}}{j} \parallel \dots\\
			&\dots \parallel \ctmthreadtw{P'_{1r}}{\ctmreturn{\!}}{t_{r}}{j} \parallel \dots
			\parallel \ctmthreadtw{P'_{2s}}{\ctmreturn{\!}}{t_{s}}{j}
		} =\ctmSt{\Sigma}{S}
		\end{align*}
		
		\begin{align*}
		\ctmSt{\Sigma}{S} \tequiv & 
		\encodingT(Q) = &\\
		=&		\encodingT(
		\tccstrana{P_{11}\mid \dots \mid P'_{1r} \mid \dots \mid P_{1m_{1}}}{C_{1}} \mid\\
		& \mid 
		\tccstrana{P_{21}\mid \dots \mid P'_{2s} \mid \dots \mid P_{2m_{2}}}{C_{2}})\\
		=&
		\ctmSt{(\Theta_{q}, \Delta_{q})}
		{
			\ctmthreadtw{\dots}{\encodingC(C_{1})}{t_{1}}{k}\parallel \ctmthreadtw{\dots}{\encodingC(C_{2})}{t_{2}}{k} \parallel \dots\\
			&\dots \parallel \ctmthreadtw{P'_{1r}}{\ctmreturn{\!}}{t_{r}}{k} \parallel \dots
			\parallel \ctmthreadtw{P'_{2s}}{\ctmreturn{\!}}{t_{s}}{k}
		}
		\end{align*}
		\item If $P  \xrightarrow{\mathsf{co} k} Q$ with \eqref{rule:tccs-trco} rule, $P = \tccstrana{R}{N}$ and $\tccstrana{R}{N} \xrightarrow{\mathsf{co} k} R'$ where $R' = \tccsprod_{i=1}^{m} R_{i}$
		\begin{align*}
		\encodingT(P) = & 
		\ctmSt{(\Theta, \Delta)}
		{
			\ctmthreadtr{\texttt{recv co } l_{t}\ \ctmreturn{\!}}{\encodingC(N)}{R}{t}{k} \parallel 
			\ctmthreadtw{\ctmreturn{t_{m}}}{\ctmreturn{\!}}{t'}{k} \parallel \\
			&\parallel
			\ctmthreadtw{\encodingC(R_{1})}{\ctmreturn{\!}}{t_{1}}{k} \parallel 
			\dots \parallel
			\ctmthreadtw{\encodingC(R_{j})}{\ctmreturn{\!}}{t_{j}}{k}
			\parallel \dots\\
			&\dots \parallel
			\ctmthreadtw{\encodingC(R_{m})}{\ctmreturn{\!}}{t_{m}}{k}
		} \\
		& \mbox{where}\\
		&\quad \Delta(l_{t})= (\texttt{True}, k),\ \Delta(\texttt{co})=(\texttt{M0},k) \\
		&\quad R = \texttt{bang psi}
		\end{align*}
		$\exists j \in \{1,\dots,m\},\ R_{j}=co.R'_{j},\ \encodingC(R_j)= \texttt{send co } l\ \encodingC(R'_{j})$.
		
		\begin{align*}
		\encodingT(P) \rightarrow^* & 
		\ctmSt{(\Theta, \Delta'')}
		{
			\ctmthreadtr{\texttt{recv co } l_{t}\ \ctmreturn{\!}}{\encodingC(N)}{R}{t}{k} \parallel 
			\ctmthreadtw{\ctmreturn{t_{m}}}{\ctmreturn{\!}}{t'}{k} \parallel \\
			&\parallel
			\dots \parallel
			\ctmthreadtw{\texttt{send co } l_{t_{j}}\ \encodingC(R'_{j})}{\ctmreturn{\!}}{t_{j}}{k}
			\parallel 
			\dots
		}
		\end{align*}	
		At this point threads $t, t_{j}$ can synchronize through $co$ variable and transaction $k$ can commit.		
		\begin{align*}
		\rightarrow^* & 
		\ctmSt{(\Theta, \Delta'')}
		{
			\ctmthreadtr{\texttt{recv co } l_{t}\ \ctmreturn{\!}}{\encodingC(N)}{\texttt{bang psi}}{t}{k} \parallel \\
			& \parallel
			\ctmthreadtw{\ctmreturn{t_{m}}}{\ctmreturn{\!}}{t'}{k} \parallel
			\ctmthreadtw{\encodingC(R_1)}{\ctmreturn{\!}}{t_{1}}{k} \parallel 
			\dots \\
			&\dots\parallel
			\ctmthreadtw{\texttt{send co } l_{t_{j}}\ \encodingC(R'_{j})}{\ctmreturn{\!}}{t_{j}}{k}\parallel \dots
			\parallel 
			\ctmthreadtw{\encodingC(R_m)}{\ctmreturn{\!}}{t_{m}}{k}
		}
		\\
		\rightarrow^* &
		\ctmSt{(\Theta, \Delta''')}
		{
			\ctmthreadtr{\ctmreturn{\!}}{\encodingC(N)}{\texttt{bang psi}}{t}{k} \parallel \\
			&\parallel 
			\ctmthreadtw{\ctmreturn{t_{m}}}{\ctmreturn{\!}}{t'}{k} \parallel \\
			&\parallel
			\ctmthreadtw{\encodingC(R_1)}{\ctmreturn{\!}}{t_{1}}{k} \parallel 
			\dots \parallel
			\ctmthreadtw{ \encodingC(R'_{j}) }{\ctmreturn{\!}}{t_{j}}{k}\parallel \dots
			\\
			&\dots \parallel 
			\ctmthreadtw{\encodingC(R_m)}{\ctmreturn{\!}}{t_{m}}{k}
		}\\
		\coArrow&
		\ctmSt{\Sigma}
		{
			\ctmthread{ \texttt{bang psi}} \parallel 
			\ctmthread{\ctmreturn{t_{m}}}{t'} \parallel \\
			&\parallel
			\ctmthread[t_{1}]{\encodingC(R_1)} \parallel 
			\dots \parallel
			\ctmthread[t_{j}]{ \encodingC(R'_{j}) }\parallel \dots \parallel 
			\ctmthread[t_{m}]{\encodingC(R_m)}
		} = \ctmSt{\Sigma}{S}\\
		& \mbox{where } \Sigma=\mathsf{commit}(k, \Theta, \Delta''')
		\end{align*}		
		$\ctmSt{\Sigma}{S} \tequiv \encodingT(Q)$,
		$ Q = R' = \tccsprod_{1}^{m}R_{i}$ and $\exists j \in \{1,\dots,m\} : R_{j} = R'_{j} $
		
		$$
		\encodingT(Q) = \encodingT(R') = \ctmSt{\Sigma_q}{
			\ctmthread[t_1]{\encodingC(R_1)}\parallel
			\dots\parallel
			\ctmthread[t_j]{\encodingC(R'_{j})}
			\parallel \dots \parallel
			\ctmthread[t_m]{\encodingC(R_m)}
		} \tequiv \ctmSt{\Sigma}{S}
		$$
	\end{enumerate}
\end{proofatend}
\begin{lemma}\label{prop:2}
	For $P\in Proc$,
	for all $S,\Sigma$, if $\encodingT(P)\rightarrow \ctmSt{\Sigma}{S}$
	then there exist $Q,S',\Sigma'$ such that
	$\ctmSt{\Sigma'}{S'}  \tequiv \encodingT(Q)$ and one of the
	following holds:		
  \begin{itemize}
  \item $P \xrightarrow{\tau}_\sigma Q$ or
    $P \xrightarrow{k(\tau)}_\sigma Q$, and
    $\ctmSt{\Sigma}{S} \rightarrow^* \ctmSt{\Sigma'}{S'}$;
  \item $P \xrightarrow{\beta}_\epsilon Q$ and
    $\ctmSt{\Sigma}{S} \xrightarrow{\smash{\beta}}^*  \ctmSt{\Sigma'}{S'}$.
  \end{itemize} 	
\end{lemma}
\begin{proofatend}
	The proof goes through induction on the semantic of \tccsm. Here we show only 3 cases, first when $P$ are two processes that can perform a synchronization outside transactions, second $P$ synchronizes inside transactional processes, third a transactional process commits.	
	\begin{enumerate}		
		\item If $P = a.P_{1} \mid \overline{a}.P_{2}$ 		
		From $\encodingT(P)$ we can move to another state of the \octm machine $\ctmSt{\Sigma}{S}$ and $\ctmSt{\Sigma}{S}\tequiv \encodingT(Q)$:
		\begin{align*}
		\encodingT(P) =&	
		\ctmSt{(\Theta',\emptyset)}{
			\ctmthread[t_1]{\ctmreturn{t'_1}} \parallel 
			\ctmthread[t'_1]{\texttt{recv}\ a\ r_{l}\ \encodingC(P_1)} \parallel \\
			&\parallel
			\ctmthread[t_2]{\ctmreturn{t'_2}} \parallel
			\ctmthread[t'_2]{\texttt{recv}\ a\ r_{l}\ \encodingC(P_{2})}}
		\\
		\rightarrow^*&
		\ctmSt{(\Theta',\emptyset)}{
			\ctmthread[t_1]{\ctmreturn{t'_1}} \parallel \ctmthread[t'_1]{\encodingC(P_{1})} \parallel
			\ctmthread[t_2]{\ctmreturn{t'_2}} \parallel \ctmthread[t'_2]{\encodingC(P_{2})}
		} = \ctmSt{\Sigma}{S}
		\\ & \mbox{where } \Theta'(l_{t_{1}}) = False,\ \Theta'(l_{t_{2}}) = False\ \Theta'(a) = \texttt{M0}&\\
		\end{align*}
		If $P \xrightarrow{\tau} Q$, then $Q = P_{1} \mid P_{2}$, $\encodingT(Q) = \ctmSt{(\Theta_{q},\emptyset)}{
			\ctmthread[t_1]{\encodingC(P_{1})} \parallel \ctmthread[t_2]{\encodingC(P_{2})}}$, we can observe $\ctmSt{\Sigma'}{S'} \tequiv \encodingT(Q)$.
		
		\item If  $P = \tccstrana[i]{a.P_{1}}{Q_{1}} \mid \tccstrana[j]{\overline{a}.P_{2}}{Q_{2}}$, $\encodingT(P) \rightarrow \ctmSt{\Sigma}{S} \rightarrow^{*} \ctmSt{\Sigma'}{S'}$ the computations are exactly the same as the previous point, but all variables are tentative in $\Delta$. It is easy to see that $P \xrightarrow{\tau(k)}_{i,j\mapsto k} Q$ and $\ctmSt{\Sigma}{S} \tequiv \encodingT(Q)$.
		
		\item If $P = \tccstrana{co.P'}{Q}$		
		\begin{align*}
		\encodingT(P) =&\\
		=& 
		\ctmSt{(\Theta, \Delta)}{
			\ctmthreadtr{\texttt{recv}\ co\ l\  \ctmreturn{\!}}{\encodingC(Q)}{\texttt{bang psi}}{t}{k} \parallel
			\\
			&	\parallel \ctmthreadtw{\texttt{send}\ co\ l\ \encodingC(P')}{\ctmreturn{\!}}{t'}{k}
		}
		\\
		\trArrow&
		\ctmSt{(\Theta, \Delta')}{
			\ctmthreadtr{M}{\encodingC(Q)}{\texttt{bang psi}}{t}{k}\parallel
			\\
			&\parallel	\ctmthreadtw{\texttt{send}\ co\ l\ \encodingC(P')}{\ctmreturn{\!}}{t'}{k}
		} = \ctmSt{\Sigma}{S}
		\\
		&\mbox{where } \Delta'(np) = nonce_{t_{1}}
		\\
		\rightarrow^{*}\coArrow&
		\ctmSt{(\Theta', \emptyset)}{
			\ctmthread[t]{\texttt{bang psi}}\parallel\ctmthread[t']{\encodingC(P')}
		}
		\\ &\mbox{where } \Sigma'=\mathsf{commit}(k, \Theta', \Delta')&
		\\ &=\ctmSt{\Sigma'}{S'}\\
		\end{align*}
		
		$P = \tccstrana{co.P'}{N}\xrightarrow{\mathsf{co}k} P' = Q$, $\encodingT(Q) = \ctmSt{\Sigma_q}{\ctmthread[t_q]{\encodingC(P')}}$, $\encodingT(Q)\tequiv \ctmSt{\Sigma'}{S'} $		
	\end{enumerate}
\end{proofatend}

\begin{theorem}
  For all $P\in Proc$, $P$ is star simulated by $\encodingT(P)$ up to
  $\tequiv$.
\end{theorem}

\section{Conclusions and future work}\label{sec:concl}
In this paper we have introduced \octm, a higher-order language
extending the concurrency model of STM Haskell with composable
\emph{open (multi-thread)} transactions. In this language, processes
can \emph{join} transactions and transactions can \emph{merge} at
runtime. These interactions are driven only by access to shared
transactional memory, and hence are implicit and loosely coupled.  To
this end, we have separated the isolation aspect from atomicity: the
\texttt{atomic} construct ensures ``all-or-nothing'' execution but not
isolation, while the new constructor \texttt{isolated} can be used to
guarantee isolation when needed.  In order to show the expressive
power of \octm, we have provided an adequate implementation in it of
\tccsm, a recently introduced model of open transactions with CCS-like
communication.  As a side result, we have given a simple typing system
for capturing \tccsm well-formed terms.

Several directions for future work stem from the present paper.
First, we plan to implement \octm along the line of STM Haskell,
but clearly the basic ideas of \octm are quite general and can be applied to other STM
implementations, like C/C++ LibCMT and Java Multiverse.  

An interesting possibility is to use \tccsm as an \emph{exogenous
  orchestration language} for \octm: the \emph{behaviour} of a
transactional distributed system can be described as a \tccsm term,
which can be translated into a \emph{skeleton} in \octm using the
encoding provided in this paper; then, the programmer has only to
``fill in the gaps''.  Thus, \tccsm can be seen as a kind of ``global
behavioural type'' for \octm.

In fact, defining a proper behavioural typing system for transactional
languages like \octm is another interesting future work.  Some
preliminary experiments have shown that \tccsm is not enough
expressive for modelling the dynamic creation of resources (locations,
threads, etc.). We think that a good candidate could be a variant of
\tccsm with local names and scope extrusions, i.e., a ``transactional
$\pi$-calculus''.

Being based on CCS, communication in \tccsm is synchronous;
however, nowadays asynchronous models play an important r\^ole (see
e.g.~actors, event-driven programming, etc.).  It may be interesting
to generalize the discussion so as to consider also this case, e.g.~by
defining an actor-based calculus with open transactions.  Such a
calculus can be quite useful also for modelling speculative reasoning
for cooperating systems \cite{ma2010:speculative,mmp:dais14,mmp:arxiv15-distemb}.
A local version of actor-based open transactions can be
implemented in \octm using lock-free data
structures (e.g., message queues) in shared transactional memory.

\FloatBarrier

\ifappendix
\clearpage
\appendix
\printproofs
\fi
\end{document}